\documentclass[aps,prl,twocolumn,superscriptaddress,amsfont,graphicx,nofootinbib,preprintnumbers]{revtex4-1}%
\UseRawInputEncoding
\usepackage{color,graphicx,epsfig,epsf}
\usepackage{ifpdf}
\usepackage{amsmath}
\usepackage{bm}
\usepackage[english]{babel}
\usepackage{graphicx}%
\usepackage{subfigure}
\usepackage{amsfonts}%
\usepackage{amssymb}
\usepackage{braket}
\usepackage{hyperref}
\usepackage{enumerate}
\usepackage[compat=1.0.0]{tikz-feynman}

\definecolor{nicered}{rgb}{0.7,0.1,0.1}
\definecolor{nicegreen}{rgb}{0.1,0.5,0.1}
\hypersetup{colorlinks,citecolor= nicegreen,linkcolor= nicered}

\usepackage{dcolumn}
\usepackage{bm}
\usepackage{slashed}

\begin{document} 
\title{Direct Detection of Spin-Dependent Sub-GeV Dark Matter via Migdal Effect}
\author{Wenyu Wang}
\email{wywang@bjut.edu.cn}
\affiliation{Faculty of Science, Beijing University of Technology, Beijing 100124, China} 

\author{Ke-Yun Wu}
\email{keyunwu@emails.bjut.edu.cn}
\affiliation{Faculty of Science, Beijing University of Technology, Beijing 100124, China} 

\author{Lei Wu}
\email{leiwu@njnu.edu.cn}
\affiliation{Department of Physics and Institute of Theoretical Physics, Nanjing Normal University, Nanjing 210023, China}

\author{Bin Zhu}
\email{zhubin@mail.nankai.edu.cn}
\affiliation{School of Physics, Yantai University, Yantai 264005, China}
\affiliation{Department of Physics, Chung-Ang University, Seoul 06974, Korea}

\begin{abstract}
Motivated by the current strong constraints on the spin-independent dark matter (DM)-nucleus scattering, we investigate the spin-dependent (SD) interactions of the light Majorana DM with the nucleus mediated by an axial-vector boson. Due to the small nucleus recoil energy, the ionization signals have now been used to probe the light dark matter particles in direct detection experiments. With the existing ionization data, we derive the exclusion limits on the SD DM-nucleus scattering through Migdal effect in the MeV-GeV DM mass range. It is found that the lower limit of the DM mass can reach about several MeVs. Due to the momentum transfer correction induced by the light mediator, the bounds on the SD DM-nucleus scattering cross sections can be weakened in comparison with the heavy mediator.

\end{abstract}
\pacs{}
\maketitle

\section{Introduction}
About 27\% of the energy density of the Universe consists of dark matter. It is non-luminous and rarely interacts with baryonic matter. In the past decades, different methods have been proposed to unveil the identity of the DM. However, only gravitational effects of the DM have been observed so far. The nature of DM is still elusive so that the detection of dark matter is one of the highest priorities in particle physics.

An attractive DM candidate is Weakly Interacting Massive Particles (WIMPs), which have been explored in various direct detection, indirect detection and collider experiments. The null results have produced very stringent limits on the WIMP-nucleus scattering cross section~(see a recent review~\cite{Roszkowski:2017nbc} and the reference therein). On the other hand, if the DM mass is approximately below $\sim 1$ GeV, it will not have enough kinetic energy to produce an observable nuclear recoil in conventional detector because of the low momentum transfer. 

Instead, the ionization signal provides a promising way to detect sub-GeV DM. For example, in a dual-phase xenon Time Projection Chamber, the xenon atoms in the Liquid Xenon phase can be ionized due to a collision. Then the ionized electrons drift into the Gaseous Xenon layer at the top of the detector in presence of an external electric field, and then collide with the xenon atoms, which produces a proportional scintillation light, namely the S2 signal. In theory, such ionization signals can come from the DM-electron scattering~\cite{Essig:2011nj,Essig:2012yx,Essig:2015cda,Chen:2015pha,Essig:2017kqs,Fornal:2020npv,Gao:2020wer,Athron:2020maw,Su:2020zny,Cao:2020bwd,Ge:2020jfn,Choi:2020ysq,An:2020bxd,Bloch:2020uzh,Zu:2020bsx,Guo:2020oum,Du:2020ybt,Knapen:2021run,Chao:2021liw,Hochberg:2021pkt} or the DM-nucleus scattering through the Migdal effect that originates from non-instantaneous movement of electron cloud during a nuclear recoil event~\cite{Migdal:1939,Vergados:2005dpd,Moustakidis:2005gx,Ejiri:2005aj,Bernabei:2007jz,Ibe:2017yqa,Dolan:2017xbu,Bell:2019egg,Essig:2019xkx,Baxter:2019pnz,GrillidiCortona:2020owp,Liu:2020pat,Knapen:2020aky,Flambaum:2020xxo,He:2020sat,Liang:2020ryg,Bell:2021zkr,Acevedo:2021kly}. The Migdal scattering is usually sub-dominant to the conventional nuclear scattering, but can take place in a very low energy nuclear recoil, which has been used to improve the sensitivity of the DM-nucleus interactions in the low DM mass region~\cite{CDEX:2019hzn,XENON:2019zpr,COSINE-100:2021poy}. 

Due to the coherent effect, it is often assumed that the spin-independent (SI) interactions are first observed in the direct detection. However, there are some exceptional that the SD interactions can be the dominant effects at the leading order, or even the only interactions accessible in direct detection experiments~\cite{Agrawal:2010fh,Fan:2010gt,Freytsis:2010ne,Fitzpatrick:2012ix,Fitzpatrick:2012ib,Anand:2013yka}, such as models with a light pseudoscalar~\cite{Freytsis:2010ne} and SI blind spots in the minimal supersymmetric standard model~\cite{Cheung:2012qy,Huang:2014xua,Han:2016qtc,Abdughani:2017dqs}. Besides, some SI interactions can also be induced by the SD couplings at the sub-leading order, which may be seen in SI direct detection simultaneously or in the next generation of experiments after the SD scattering is observed. On the other hand, the detection of the SD interactions could be used to identify the spin of the DM. For example, the scalar DM can only induce the SI scattering and thus would be excluded if any SD signal is observed in the direct detection experiments.


In this work, we focus on the SD interactions of the DM and nucleus mediated by a vector boson in the simplified model. We calculate the event rates of the SD DM-nucleus scattering through the Migdal effect. Since the light mediator can induce the sizable momentum-dependence correction~\cite{Li:2014vza,Ramani:2019jam}, we also include such effects in our calculations. By utilizing the ionization data of XENON10/100/1T, we derive the constraints on the SD interactions of DM-nucleus in the MeV-GeV DM mass range.

\section{SD DM-nucleus Scattering through Migdal Effect}\label{sec2}

In theories where the DM couples predominantly to the spin of the nucleus, the corresponding interactions are dubbed as spin-dependent. In contrast with the SI couplings, there is no coherent enhancement in the spin-dependent DM-nucleus scattering cross section because the spins of nucleons in a nucleus tend to cancel in pairs. The inputs of the calculations of the SD DM-nucleus scattering include the type of DM-nucleus coupling as well as the nuclear response to the DM interaction, which rely on the new physics inducing the scattering and known nuclear physics, respectively. We assume a Majorana fermion DM ($\chi$) and an axial-vector mediator ($A_\mu$) and perform the calculations in a simplified model. The interactions of the mediator with the DM and quarks are given by,
\begin{eqnarray}
 {\cal L} \supset  g_\chi \bar{\chi}\gamma_\mu \gamma_5 \chi A^\mu + g_q \bar{\psi}_q \gamma_\mu \gamma_5 \psi_q A^\mu \label{lag}
\end{eqnarray} 
This leads to a spin-spin interaction $-4 {\bm S}_\chi \cdot {\bm S}_N$ in the non-relativistic effective field theory~\cite{Agrawal:2010fh,Fan:2010gt,Freytsis:2010ne,Fitzpatrick:2012ix}. 
For the Majorana fermion DM, such an interaction is the solely SD coupling that is not suppressed by the DM velocity at the leading order. 

If the mediator is heavy, the effective Lagrangian of the DM interactions can be written as~\cite{Engel:1992bf,Jungman:1995df},
\begin{eqnarray}
{\mathcal L}^{\rm SD}_\chi &=& \frac{G_{v}}{\sqrt{2}}
\int d^4{x} \, {j}^\mu({x}) {J}^A_\mu({x})\label{relaL}
\end{eqnarray}
where ${J}^A_\mu({x})= \bar{\psi}_q { \gamma}_\mu \gamma_5 \psi_q$ and ${j}^\mu({x})=\bar\chi { \gamma}^\mu \gamma^5 \chi$ denote the hadronic current and the DM current, respectively. $G_v=\sqrt{2}g_\chi g_q/m^2_A$ is the coupling of the DM and quarks. From Eq.~\eqref{relaL}, we can obtain the structure factor $S_N(q)$ that represents the spin structure of the nucleus currents in the target, At $q=0$,
\begin{eqnarray}
    S_N(0)&=&\frac{(2J+1)(J+1)}{4\pi J}\\
    && \times\left|(a_0+a_1)\langle{\bm S}_p\rangle+
    (a_0-a_1)\langle{\bm S}_n\rangle\right|^2,\nonumber
\end{eqnarray}
where $J$ is the total angular momentum of the nucleus in a ground state. $\langle{\bm S}_p\rangle$ and $\langle{\bm S}_n\rangle$ are the expectation values of the total spin operators of proton and neutron in the nucleus, respectively. The information of the particle and hadronic sectors is encoded in the couplings $a_0$ and $a_1$. At higher $q$, due to the interference effect and $q$ dependence, the structure factor $S_N(q)$ is given by
\begin{eqnarray}
S_{N}(q)=a_0^2 S_{00}(q)+a_0a_1S_{01}(q)+a_1^2 S_{11}(q).
\end{eqnarray} 
where the functions $S_{00}$, $S_{01}$ and $S_{11}$ can be calculated from the transverse and the longitudinal electric projections of the axial current~\cite{Jungman:1995df}. By constructing the consistent WIMP-nucleon currents at the one-body level and including effects from axial-vector two-body currents, the recent results of the structure factors based on the chiral effective field theory are given in Ref.~\cite{Hu:2021awl}. Other structure functions with theoretical error bands due to the nuclear uncertainties of WIMP currents in nuclei are provided in Ref.~\cite{Klos:2013rwa}. Besides, as shown in Ref.~\cite{Wang:2021nbf}, when the DM moves fast, the structure factor $S_N(q)$ is not only a function of momentum transfer $q$, but also the incoming DM energy of $E_\chi$. Such an effect may weaken the SD exclusion limits.  

Since the momentum transfer in DM-nucleus scattering is generally smaller than the characteristic nuclear scale, we define the reference scattering cross section at zero momentum transfer~\cite{Bednyakov:2004xq,Bednyakov:2006ux}, $\bar{\sigma}^{\rm SD}_{\chi N}$, as
\begin{eqnarray}
\bar{\sigma}^{\rm SD}_{\chi N} = \int  \frac{d\sigma^{\mathrm{SD}}_{\chi N}}{dq^2}(q^2 = 0) dq^2,
\label{sigmaN}
\end{eqnarray}
On the other hand, if the mediator is light, the momentum transfer effect can be sizable. To include this correction, we define the DM form factor $F_{\rm DM}(q)$ as
\begin{eqnarray}
F_{\rm DM}(q) \equiv \frac{1}{1+q^2/m_A^{2}}.
\end{eqnarray}
Thus, we can have the full differential SD DM-nucleus scattering cross section,
\begin{eqnarray}
\frac{d\sigma^{\mathrm{SD}}_{\chi N}}{dq^2} = 
\frac{\bar{\sigma}^{\rm SD}_{\chi N}}{3\mu_{\chi N}^2 v^2 }
\frac{\pi }{2J_N+1}S_N(q)  |F_{\rm DM}(q)|^2.
\label{sigmaN}
\end{eqnarray}

In order to obtain the ionization events, we also need to calculate the possibility of the electrons ionized by Migdal scattering. We assume $\vert i\rangle$ is the electron cloud state in the rest frame of the nucleus and $\vert f \rangle$ is the finial ionization state of atom. The transition rate from $\vert i\rangle$ to $\vert f \rangle$ is given by,
\begin{eqnarray}
\langle f\vert e^{-i\frac{m_e}{m_A}\bm{q}\cdot\sum_{s}\bm{x}^{s}}\vert i \rangle\approx\langle f\vert e^{-i\frac{m_e}{m_N}\bm{q}\cdot\sum_{s}\bm{x}^{s}}\vert i \rangle, 
\end{eqnarray}
where $m_A$, $m_{e}$,  $m_N$ are the masses of the atom, the electron and the nucleus, respectively. The position of $s^{th}$ electron is denoted by $\bm {x}^s$ in different atomic shells. The $\bm {q}$ is the three momentum transfer for the DM-nucleus scattering. While for the ionized electron, the momentum transfer is $\bm{q}_{e}\simeq m_{e}\bm{q}/m_{N}$.
Since the ionization factor $\vert f^{ion}_{nl}(p_e ,q_e)\vert^2$ is proportional to the ionization possibility, we can express it as,
\begin{eqnarray}
\vert f_{\rm ion, M}^{nl}(k_e,q_e)\vert^{2} &=& \frac{4k^{3}_e}{(2\pi)^{3}}
	\sum_{n,l,l',m'}\vert \langle f\vert e^{i \bm{q_{e}}\cdot\bm{x}_i}\vert i\rangle\vert^{2},
	\label{fion}
\end{eqnarray}
in which, $k_{e}=\sqrt{2E_{R}m_{e}}$. In our calculation, we include the contributions from the electron configurations of $4s$, $4p$, $4d$, $5s$ and $5p$, and compute the ionization factor by numerically solving Schr{\"o}dinger equation with the package \textsf{DarkARC}~\cite{Catena:2019gfa}. Since the momentum transfer to the electron is small, the relativistic correction to the ionization factor is negligible~\cite{Roberts:2016xfw}. In numerical calculation, when $q_e$ is very small, we note that the orthogonality of overlap wave functions can be violated due to numerical stability, and thus adopt the subtraction method in the Ref.~\cite{Tan:2021nif}.

Finally, the differential event rate of the DM scattering off a nucleus in a ground state with total angular momentum $J_N$ through the Migdal effect is given by,
\begin{eqnarray}
\frac{d R^{\rm ion}_{M}}{d \ln E_{R}} &=& N_T\frac{\rho_{\chi}}{m_{\chi}}
\frac{\bar{\sigma}^{\rm SD}_{\chi N}}{3\mu_{\chi {N}}^{2}}\frac{\pi }{2J_N+1} \sum_{n,l} \int^{q_{\rm max}}_{q_{\rm min}}  q dq  \\
&& \times \eta(v_{\rm min}) S_{N}(q) |F_{DM}(q)|^2 |f^{ion}_{nl}(k_e,q_e)|^2,\nonumber
\label{drate}
\end{eqnarray}
where $N_T$ is the number of the target atom and $\rho_{\chi}=0.4~\rm GeV/cm^{3}$ is the local DM density. $\mu_{\chi N}=m_\chi m_N/(m_\chi + m_N)$ is the reduced mass of DM-nucleus system and $(n,l)$ denotes the atomic shells. $q$ is the amplitude of the momentum transferred by the DM to the nucleus. The inverse mean speed function $\eta (v_{\rm min})$ for the normal Maxwell-Boltzmann velocity distribution with the circular velocity $v_0=220$ km/s is given by
\begin{eqnarray}
  \eta(v_{\rm min})=\int_{v_{\rm min}} \frac{d^3 v}{v} f_\chi(v) \Theta (v-v_{\rm min}),
\end{eqnarray}
with
\begin{eqnarray}
 f_\chi(\vec{v}_\chi) \propto e^{-\frac{|\vec{v}_\chi +\vec{v}_E|^2}{v^2_0}} \Theta (v_{\rm esc}-|\vec{v}_{\chi}+\vec{v}_E|),
\end{eqnarray}
where we take the escape velocity $v_{\rm esc}=544$ km/s and the averaged Earth relative velocity $v_E=232$ km/s~\cite{Smith:2006ym,Dehnen:1997cq} so that $v_{\rm max}=v_{\rm esc}+v_E$. Since the recoil energy of the nucleus is negligible, the energy transferred to the electron $\Delta E_e$ is given by
\begin{equation}
\Delta E_e= {\bm v} \cdot {\bm q} - \frac{{\bm q}^2}{2m_\chi}.
\label{dee}
\end{equation}
The recoil energy of the electron depends on the binding energy of its state so that $\Delta E_e = E_{R}+E_b$, where $E_b$ is the binding energy of the electron and $E_{R}$ is the recoil energy of the electron. Then, we can have the minimum velocity $v_{\rm min}$ required for ionizing the electron through the momentum transfer $q$ in the DM-nucleus Migdal scattering,
\begin{equation}
v_{\rm min}=\frac{|E_b|+E_{R}}{q}+\frac{q}{2\mu_{\chi N}}.
\label{vmin}
\end{equation}
Here the minimum $q_{\rm min}$ and maximum $q_{\rm max}$ in Eq.~\eqref{vmin} are determined by
\begin{equation}
q_{\rm min}=\frac{|E_{b}|}{v_{\rm max}},\quad q_{\rm max}=2\mu_{\chi N}v_{\rm max}.
\label{qm}
\end{equation} 
Besides, from Eq.~\eqref{vmin}, we can derive the maximum electron recoil energy $E^{\rm max}_{R}$,
\begin{equation}
E^{\rm max}_{R}=\frac{1}{2}\mu_{\chi N}v^2_{\rm max},
\end{equation} 
which is much larger than the maximum nucleus recoil $E^{\rm max}_{nr}=2 \mu^2_{\chi N} v^2_{\rm max}/m_N$ for a sub-GeV DM.

\section{Numerical Results and Discussions}\label{sec3}

\begin{figure}[ht]
\centering
\epsfig{file=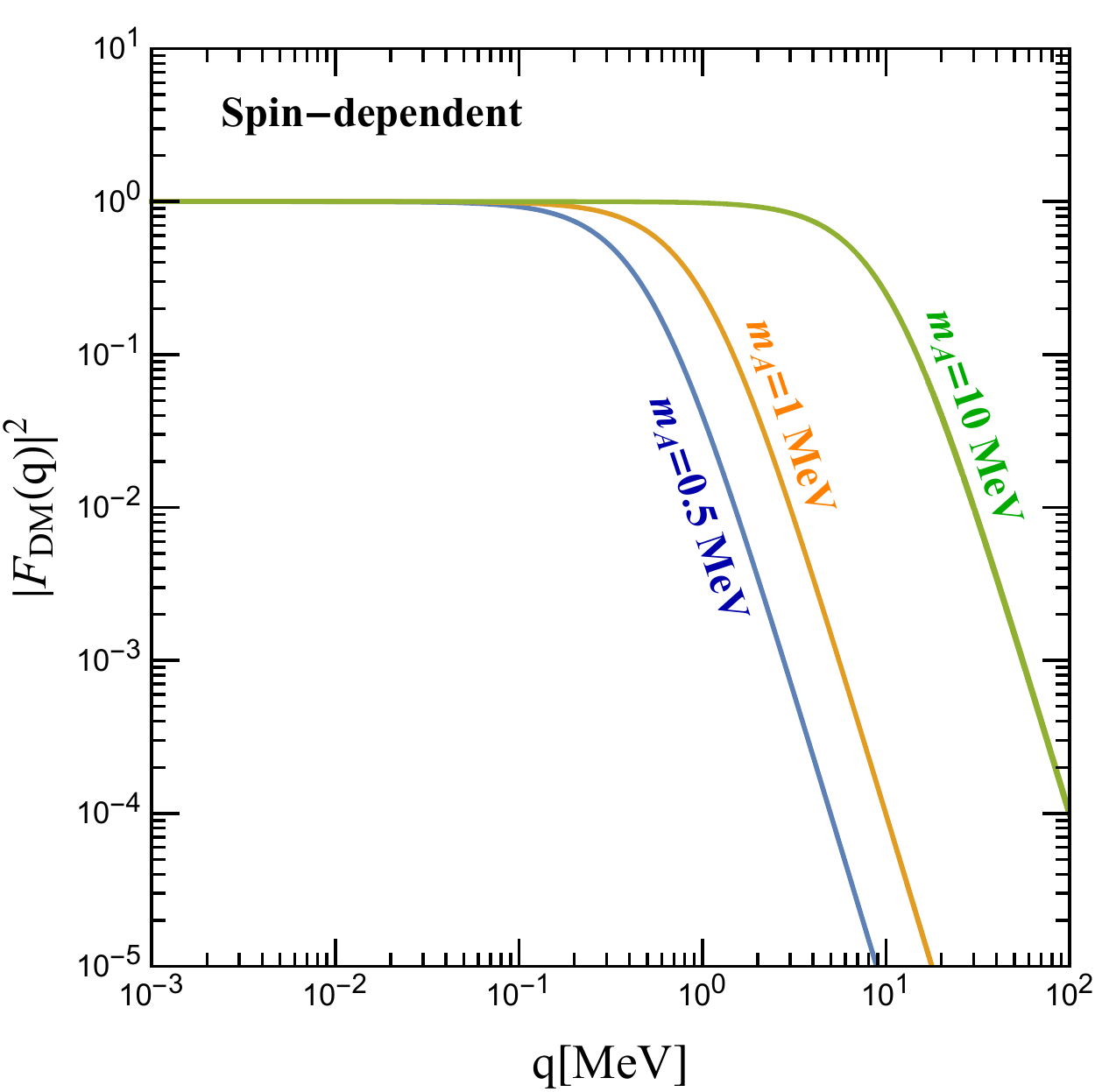, width=8cm, height =8cm}
\caption{The dependence of the DM form factor $|{\rm F_{DM}}(q)|^{2}$ on the momentum transfer $q$ for the mediator mass $m_A = 0.5,~1,~10$ MeV. }
\label{fdm}	
\end{figure}
In Fig.~\ref{fdm}, we show the dependence of $|F_{\rm DM}(q)|^2$ on the momentum transfer $q$ for the mediator masses $m_A=0.5~,1~,10$ MeV. It can be seen that the DM form factor is highly suppressed with the increase of $q$. For a heavier mediator, the suppression effect appears in a higher $q$ region. This will affect the bounds on the DM-nucleus scattering cross section in different DM mass range.

Following the procedure in Ref.~\cite{Essig:2012yx}, we derive the constraints on the SD scattering cross section from the experimental data. The nucleus recoil $E_R$ can induce both of the observed electrons $n_{e}$ and scintillation photons $n_{\gamma}$. The numbers of quanta $n^{(1)}$ that are produced by the step energy $W=13.8$~eV from the nucleus recoils can be calculated by $n^{(1)} = {\rm Floor}(E_{\rm R}/W)$. The probability of the initial electron recombining with an ion is assumed as $f_R=0$. The fraction of the primary quanta observed as the electrons is taken as $f_e = 0.83$.  
The corresponding uncertainties are chosen as $0<f_{R}<0.2$, $12.4<W<16$ eV, and $0.62<f_{e}<0.91$.
The photons are assumed from the deexcitation of the next-to-outer shells $(5s,~4d,~4p,~4s)$, whose energies correspond to $(13.3,~63.2,~87.9,~201.4)$ eV. Due to the photoionization effect, these photons can respectively create an additional quanta numbers $n^{(2)} = (n_{5s},~n_{4d},~n_{4p},~n_{4s}) = (0,~4,~6-10,~3-15)$ as shown in the Table~\ref{tabqe}
\cite{Essig:2017kqs}. The number of electrons and photons in the total quanta ($n^{(1)}+n^{(2)}$) 
obey a binomial distribution. Then, we can obtain the electron event $n_{e}$ according to the distribution. In experiments, the ionized electrons will be converted into the photoelectron (PE). The number of PE can be produced by an event of 
$n_e$ which obey gaussian distribution with mean $n_{e}\mu$ and width $\sqrt{n_{e}\sigma}$, 
where $\mu = 27~(19.7, 11.4)$ and $\sigma = 6.7~ (6.2, 2.8)$ for XENON10 (XENON100, XENON1T). 
The corresponding bins can be found in
Ref.~\cite{Essig:2012yx,XENON100:2011cza,XENON:2019gfn}. The event rate of PE from XENON10 data \cite{Essig:2012yx} (15 kg-days), 
XENON100 data \cite{XENON100:2011cza, XENON:2016jmt} (30 kg-years) and 
XENON1T data \cite{XENON:2019gfn} (1.5 tones-years) are used to constrain our SD scattering cross 
section.

\begin{table}[ht]
	\begin{tabular}{| l || l | l | l | l | l |}
		\hline 
		shell& 5$p^{6}$ & 5$s^{2}$ & 4$d^{10}$ & 4$p^{6}$ & 4 $s^{2}$\\
		\cline{1-6}
		Binding Energy[eV] & 12.6 & 25.7 & 75.6 & 163.5 & 213.8\\	
		\cline{1-6}
		Additional Quanta & 0 & 0 & 4 & 6-10 & 3-15\\
		\cline{1-6}
		\hline
	\end{tabular}
\caption{The number of additional quanta and binding energy 
from the Xenon ($5p^{6},~5s^{2},~4d^{10},~4p^{6},~4s^{2}$) shells. }\label{tabqe}
\end{table}

\begin{figure}[ht]
\centering
\epsfig{file=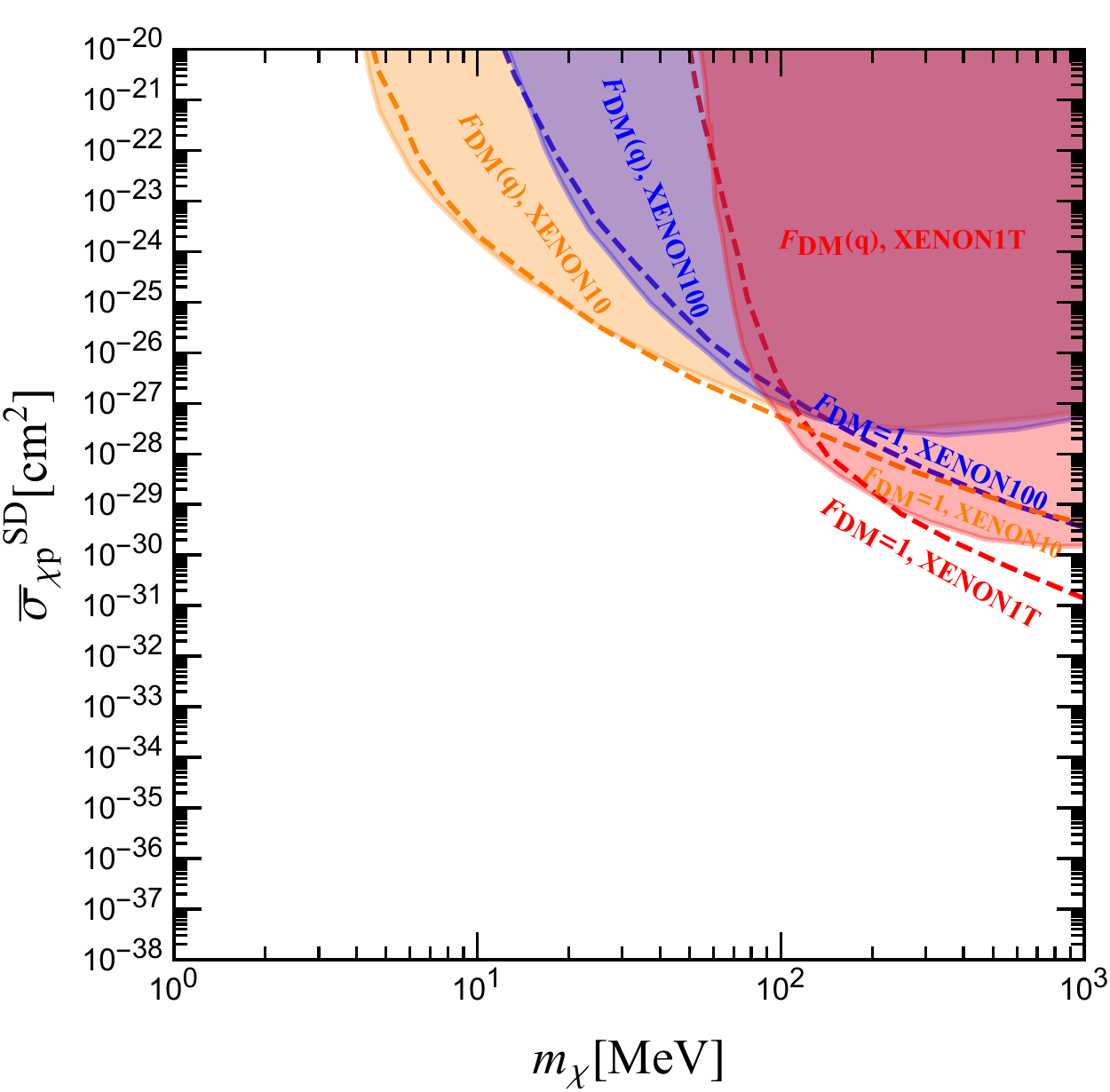, width=8cm, height =8cm}
\epsfig{file=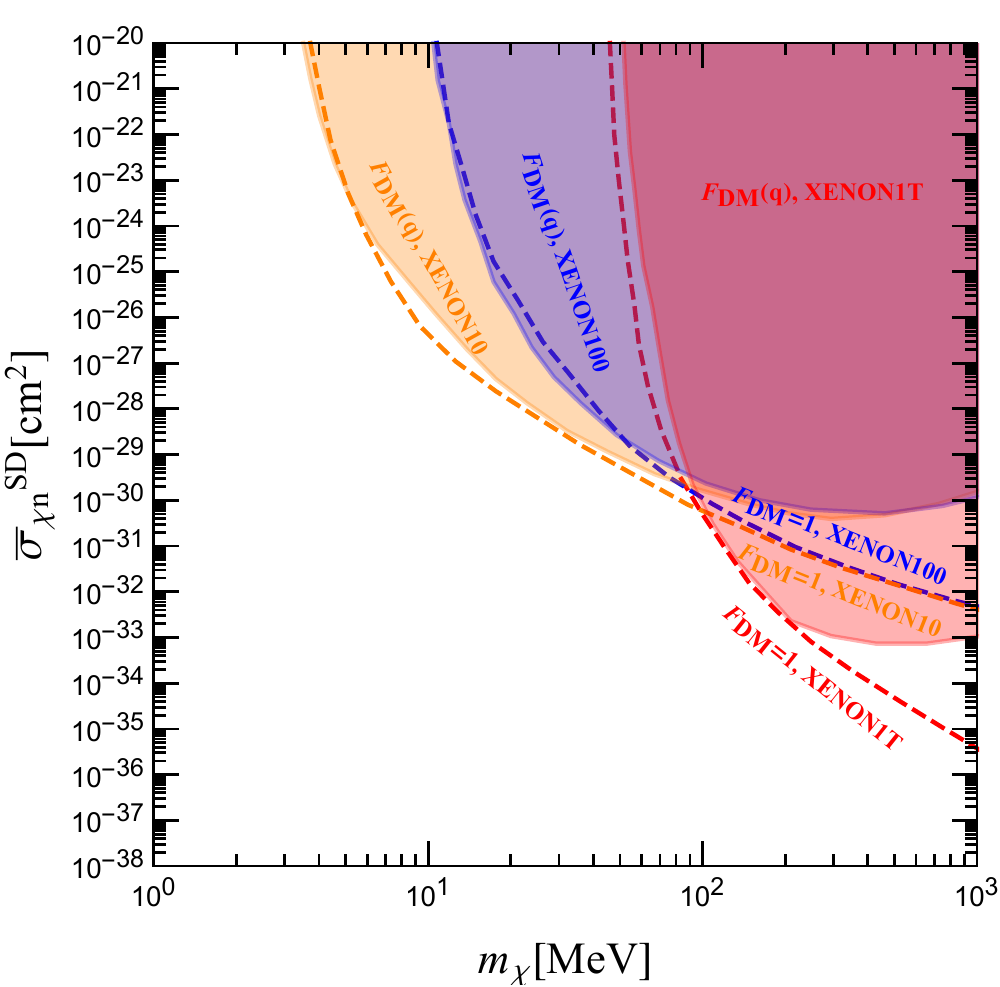, width=8cm, height =8cm}
\caption{The exclusion limits for the SD DM-proton (top) and DM-neutron (bottom) scattering cross sections from XENON10 (orange), XENON100 (blue) and XENON1T (red) data. The mediator mass $m_A=0.5$ MeV is assumed in the calculations.  As a comparison, the results for $F_{\rm DM}=1$ (dotted lines) are also shown in each panel.}	\label{sigmap}	
\end{figure} 

The limits of reduced SD scattering cross sections $\bar{\sigma}_{\chi p}^{SD}$ (for the proton) and $\bar{\sigma}_{\chi n}^{SD}$ (for the neutron) are shown
in the Fig.~\ref{sigmap}, where we take the light mediator mass $m_A=0.5$ MeV as an example. As a comparison, the results for the heavy mediator with $F_{DM}=1$ are also presented in the dash lines. From Fig.~\ref{sigmap}, we can see that the exclusion upper limit of the SD DM-neutron scattering cross section derived from the XENON1T data can reach $\bar{\sigma}^{\rm SD}_{\chi n} \sim 10^{-33}$ cm$^2$ at $m_\chi=500$ MeV, which is about three orders of magnitude stronger than those on the DM-proton scattering cross section in both light and heavy mediator cases. This is because that the structure factor $S_n(q)$ of the neutron in Eq.~\ref{drate} are much larger than $S_p(q)$ of the proton. 

Besides, we find that the bounds for the light and heavy mediators are almost the same when the DM mass is below about 100 MeV. On the other hand, since the light mediator can induce a $\sim 1/q^4$ suppression, the exclusion limits for the light mediator becomes weak with the increase of the DM mass, which can be about two orders of magnitude smaller than those for the heavy mediator. We can understand such a feature from the Fig.~\ref{fdm}. For $m_A=0.5$ MeV, the DM form factor begins to decrease from about $q=0.1$ MeV. Then, with Eq.~\ref{qm}, we can estimate that the $q_{\rm max} \sim 0.1$ MeV at $m_\chi=100$ MeV. This indicates that the DM form factor $|F_{DM}|^2=1$ in the integral range of Eq.~\ref{drate} when the DM mass is less than 100 MeV. In other words, if the DM mass becomes heavier, the momentum transfer $q$ can be larger. In this case, although the ionization factor $f_{\rm ion}$ will increase, the DM form factor $F_{\rm DM}$ will greatly decrease and dominate the event rate of Migdal scattering. Similarly, if the light DM is boosted by some physical mechanisms, the Migdal events will be suppressed by the DM form factor as well.


Finally, it should be noted that the Fig.~\ref{sigmap} shows that the constraints from XENON10/100 data are stronger than that from XENON1T data in the region of $m_\chi < 100$ MeV. However, when $m_\chi > 100$ MeV, XENON1T data can give more stringent constraints than the other two data. The reason for this is that XENON10/100 can achieve a lower threshold of the ionization signal. With these data, the DM can be excluded down to about 4 MeV in the SD DM-nucleus scattering through the Migdal effect. Besides, like the SI scattering~(see e.g. \cite{Xia:2021vbz} and the references therein), when the DM-nucleus interaction is strong, we should mention that the earth stopping effect may attenuate the flux and produce the bound on the SD scattering cross section as well. Such a dedicated calculation will be done in our future work.


\section{Conclusion}\label{sec4}
In this work, we calculate the event rate of the spin-dependent Majorana dark matter-nucleus scattering through the Migdal effect in the simplified model with an axial-vector mediator. The constraints on the SD interactions of the DM with the nucleus for a light and heavy mediator in the MeV-GeV DM mass range are derived from the existing direct detection data. We find that the lower limit of the DM mass can reach about several MeV and the upper limit of the SD DM-neutron/proton scattering cross section is about $10^{-33}/10^{-30}$ cm$^2$ when the DM mass $m_{\chi}=500$ MeV and the mediator mass $m_{A}=0.5$ MeV. Compared with the heavy mediator, the momentum transfer effect can give a sizable correction to the result. With the decrease of the mediator mass, the exclusion limit of the SD scattering cross section will be further weakened.

\section{Acknowledgements}
We thank Liangliang Su for his helpful discussions. This work was supported by the Natural Science Foundation of China under grant number 11775012 and 11805161. The work of BZ is also supported partially by Korea Research Fellowship Program through the National Research Foundation of Korea (NRF) funded by the Ministry of Science and ICT (2019H1D3A1A01070937).

\bibliography{refs}

\begin{thebibliography}{70}%
\makeatletter
\providecommand \@ifxundefined [1]{%
 \@ifx{#1\undefined}
}%
\providecommand \@ifnum [1]{%
 \ifnum #1\expandafter \@firstoftwo
 \else \expandafter \@secondoftwo
 \fi
}%
\providecommand \@ifx [1]{%
 \ifx #1\expandafter \@firstoftwo
 \else \expandafter \@secondoftwo
 \fi
}%
\providecommand \natexlab [1]{#1}%
\providecommand \enquote  [1]{``#1''}%
\providecommand \bibnamefont  [1]{#1}%
\providecommand \bibfnamefont [1]{#1}%
\providecommand \citenamefont [1]{#1}%
\providecommand \href@noop [0]{\@secondoftwo}%
\providecommand \href [0]{\begingroup \@sanitize@url \@href}%
\providecommand \@href[1]{\@@startlink{#1}\@@href}%
\providecommand \@@href[1]{\endgroup#1\@@endlink}%
\providecommand \@sanitize@url [0]{\catcode `\\12\catcode `\$12\catcode
  `\&12\catcode `\#12\catcode `\^12\catcode `\_12\catcode `\%12\relax}%
\providecommand \@@startlink[1]{}%
\providecommand \@@endlink[0]{}%
\providecommand \url  [0]{\begingroup\@sanitize@url \@url }%
\providecommand \@url [1]{\endgroup\@href {#1}{\urlprefix }}%
\providecommand \urlprefix  [0]{URL }%
\providecommand \Eprint [0]{\href }%
\providecommand \doibase [0]{http://dx.doi.org/}%
\providecommand \selectlanguage [0]{\@gobble}%
\providecommand \bibinfo  [0]{\@secondoftwo}%
\providecommand \bibfield  [0]{\@secondoftwo}%
\providecommand \translation [1]{[#1]}%
\providecommand \BibitemOpen [0]{}%
\providecommand \bibitemStop [0]{}%
\providecommand \bibitemNoStop [0]{.\EOS\space}%
\providecommand \EOS [0]{\spacefactor3000\relax}%
\providecommand \BibitemShut  [1]{\csname bibitem#1\endcsname}%
\let\auto@bib@innerbib\@empty
\bibitem [{\citenamefont {Roszkowski}\ \emph {et~al.}(2018)\citenamefont
  {Roszkowski}, \citenamefont {Sessolo},\ and\ \citenamefont
  {Trojanowski}}]{Roszkowski:2017nbc}%
  \BibitemOpen
  \bibfield  {author} {\bibinfo {author} {\bibfnamefont {L.}~\bibnamefont
  {Roszkowski}}, \bibinfo {author} {\bibfnamefont {E.~M.}\ \bibnamefont
  {Sessolo}}, \ and\ \bibinfo {author} {\bibfnamefont {S.}~\bibnamefont
  {Trojanowski}},\ }\href {\doibase 10.1088/1361-6633/aab913} {\bibfield
  {journal} {\bibinfo  {journal} {Rept. Prog. Phys.}\ }\textbf {\bibinfo
  {volume} {81}},\ \bibinfo {pages} {066201} (\bibinfo {year} {2018})},\
  \Eprint {http://arxiv.org/abs/1707.06277} {arXiv:1707.06277 [hep-ph]}
  \BibitemShut {NoStop}%
\bibitem [{\citenamefont {Essig}\ \emph
  {et~al.}(2012{\natexlab{a}})\citenamefont {Essig}, \citenamefont {Mardon},\
  and\ \citenamefont {Volansky}}]{Essig:2011nj}%
  \BibitemOpen
  \bibfield  {author} {\bibinfo {author} {\bibfnamefont {R.}~\bibnamefont
  {Essig}}, \bibinfo {author} {\bibfnamefont {J.}~\bibnamefont {Mardon}}, \
  and\ \bibinfo {author} {\bibfnamefont {T.}~\bibnamefont {Volansky}},\ }\href
  {\doibase 10.1103/PhysRevD.85.076007} {\bibfield  {journal} {\bibinfo
  {journal} {Phys. Rev. D}\ }\textbf {\bibinfo {volume} {85}},\ \bibinfo
  {pages} {076007} (\bibinfo {year} {2012}{\natexlab{a}})},\ \Eprint
  {http://arxiv.org/abs/1108.5383} {arXiv:1108.5383 [hep-ph]} \BibitemShut
  {NoStop}%
\bibitem [{\citenamefont {Essig}\ \emph
  {et~al.}(2012{\natexlab{b}})\citenamefont {Essig}, \citenamefont
  {Manalaysay}, \citenamefont {Mardon}, \citenamefont {Sorensen},\ and\
  \citenamefont {Volansky}}]{Essig:2012yx}%
  \BibitemOpen
  \bibfield  {author} {\bibinfo {author} {\bibfnamefont {R.}~\bibnamefont
  {Essig}}, \bibinfo {author} {\bibfnamefont {A.}~\bibnamefont {Manalaysay}},
  \bibinfo {author} {\bibfnamefont {J.}~\bibnamefont {Mardon}}, \bibinfo
  {author} {\bibfnamefont {P.}~\bibnamefont {Sorensen}}, \ and\ \bibinfo
  {author} {\bibfnamefont {T.}~\bibnamefont {Volansky}},\ }\href {\doibase
  10.1103/PhysRevLett.109.021301} {\bibfield  {journal} {\bibinfo  {journal}
  {Phys. Rev. Lett.}\ }\textbf {\bibinfo {volume} {109}},\ \bibinfo {pages}
  {021301} (\bibinfo {year} {2012}{\natexlab{b}})},\ \Eprint
  {http://arxiv.org/abs/1206.2644} {arXiv:1206.2644 [astro-ph.CO]} \BibitemShut
  {NoStop}%
\bibitem [{\citenamefont {Essig}\ \emph {et~al.}(2016)\citenamefont {Essig},
  \citenamefont {Fernandez-Serra}, \citenamefont {Mardon}, \citenamefont
  {Soto}, \citenamefont {Volansky},\ and\ \citenamefont {Yu}}]{Essig:2015cda}%
  \BibitemOpen
  \bibfield  {author} {\bibinfo {author} {\bibfnamefont {R.}~\bibnamefont
  {Essig}}, \bibinfo {author} {\bibfnamefont {M.}~\bibnamefont
  {Fernandez-Serra}}, \bibinfo {author} {\bibfnamefont {J.}~\bibnamefont
  {Mardon}}, \bibinfo {author} {\bibfnamefont {A.}~\bibnamefont {Soto}},
  \bibinfo {author} {\bibfnamefont {T.}~\bibnamefont {Volansky}}, \ and\
  \bibinfo {author} {\bibfnamefont {T.-T.}\ \bibnamefont {Yu}},\ }\href
  {\doibase 10.1007/JHEP05(2016)046} {\bibfield  {journal} {\bibinfo  {journal}
  {JHEP}\ }\textbf {\bibinfo {volume} {05}},\ \bibinfo {pages} {046} (\bibinfo
  {year} {2016})},\ \Eprint {http://arxiv.org/abs/1509.01598} {arXiv:1509.01598
  [hep-ph]} \BibitemShut {NoStop}%
\bibitem [{\citenamefont {Chen}\ \emph {et~al.}(2015)\citenamefont {Chen},
  \citenamefont {Chi}, \citenamefont {Liu}, \citenamefont {Wu},\ and\
  \citenamefont {Wu}}]{Chen:2015pha}%
  \BibitemOpen
  \bibfield  {author} {\bibinfo {author} {\bibfnamefont {J.-W.}\ \bibnamefont
  {Chen}}, \bibinfo {author} {\bibfnamefont {H.-C.}\ \bibnamefont {Chi}},
  \bibinfo {author} {\bibfnamefont {C.~P.}\ \bibnamefont {Liu}}, \bibinfo
  {author} {\bibfnamefont {C.-L.}\ \bibnamefont {Wu}}, \ and\ \bibinfo {author}
  {\bibfnamefont {C.-P.}\ \bibnamefont {Wu}},\ }\href {\doibase
  10.1103/PhysRevD.92.096013} {\bibfield  {journal} {\bibinfo  {journal} {Phys.
  Rev. D}\ }\textbf {\bibinfo {volume} {92}},\ \bibinfo {pages} {096013}
  (\bibinfo {year} {2015})},\ \Eprint {http://arxiv.org/abs/1508.03508}
  {arXiv:1508.03508 [hep-ph]} \BibitemShut {NoStop}%
\bibitem [{\citenamefont {Essig}\ \emph {et~al.}(2017)\citenamefont {Essig},
  \citenamefont {Volansky},\ and\ \citenamefont {Yu}}]{Essig:2017kqs}%
  \BibitemOpen
  \bibfield  {author} {\bibinfo {author} {\bibfnamefont {R.}~\bibnamefont
  {Essig}}, \bibinfo {author} {\bibfnamefont {T.}~\bibnamefont {Volansky}}, \
  and\ \bibinfo {author} {\bibfnamefont {T.-T.}\ \bibnamefont {Yu}},\ }\href
  {\doibase 10.1103/PhysRevD.96.043017} {\bibfield  {journal} {\bibinfo
  {journal} {Phys. Rev. D}\ }\textbf {\bibinfo {volume} {96}},\ \bibinfo
  {pages} {043017} (\bibinfo {year} {2017})},\ \Eprint
  {http://arxiv.org/abs/1703.00910} {arXiv:1703.00910 [hep-ph]} \BibitemShut
  {NoStop}%
\bibitem [{\citenamefont {Fornal}\ \emph {et~al.}(2020)\citenamefont {Fornal},
  \citenamefont {Sandick}, \citenamefont {Shu}, \citenamefont {Su},\ and\
  \citenamefont {Zhao}}]{Fornal:2020npv}%
  \BibitemOpen
  \bibfield  {author} {\bibinfo {author} {\bibfnamefont {B.}~\bibnamefont
  {Fornal}}, \bibinfo {author} {\bibfnamefont {P.}~\bibnamefont {Sandick}},
  \bibinfo {author} {\bibfnamefont {J.}~\bibnamefont {Shu}}, \bibinfo {author}
  {\bibfnamefont {M.}~\bibnamefont {Su}}, \ and\ \bibinfo {author}
  {\bibfnamefont {Y.}~\bibnamefont {Zhao}},\ }\href {\doibase
  10.1103/PhysRevLett.125.161804} {\bibfield  {journal} {\bibinfo  {journal}
  {Phys. Rev. Lett.}\ }\textbf {\bibinfo {volume} {125}},\ \bibinfo {pages}
  {161804} (\bibinfo {year} {2020})},\ \Eprint
  {http://arxiv.org/abs/2006.11264} {arXiv:2006.11264 [hep-ph]} \BibitemShut
  {NoStop}%
\bibitem [{\citenamefont {Gao}\ \emph {et~al.}(2020)\citenamefont {Gao},
  \citenamefont {Liu}, \citenamefont {Wang}, \citenamefont {Wang},
  \citenamefont {Xue},\ and\ \citenamefont {Zhong}}]{Gao:2020wer}%
  \BibitemOpen
  \bibfield  {author} {\bibinfo {author} {\bibfnamefont {C.}~\bibnamefont
  {Gao}}, \bibinfo {author} {\bibfnamefont {J.}~\bibnamefont {Liu}}, \bibinfo
  {author} {\bibfnamefont {L.-T.}\ \bibnamefont {Wang}}, \bibinfo {author}
  {\bibfnamefont {X.-P.}\ \bibnamefont {Wang}}, \bibinfo {author}
  {\bibfnamefont {W.}~\bibnamefont {Xue}}, \ and\ \bibinfo {author}
  {\bibfnamefont {Y.-M.}\ \bibnamefont {Zhong}},\ }\href {\doibase
  10.1103/PhysRevLett.125.131806} {\bibfield  {journal} {\bibinfo  {journal}
  {Phys. Rev. Lett.}\ }\textbf {\bibinfo {volume} {125}},\ \bibinfo {pages}
  {131806} (\bibinfo {year} {2020})},\ \Eprint
  {http://arxiv.org/abs/2006.14598} {arXiv:2006.14598 [hep-ph]} \BibitemShut
  {NoStop}%
\bibitem [{\citenamefont {Athron}\ \emph {et~al.}(2021)\citenamefont {Athron}
  \emph {et~al.}}]{Athron:2020maw}%
  \BibitemOpen
  \bibfield  {author} {\bibinfo {author} {\bibfnamefont {P.}~\bibnamefont
  {Athron}} \emph {et~al.},\ }\href {\doibase 10.1007/JHEP05(2021)159}
  {\bibfield  {journal} {\bibinfo  {journal} {JHEP}\ }\textbf {\bibinfo
  {volume} {05}},\ \bibinfo {pages} {159} (\bibinfo {year} {2021})},\ \Eprint
  {http://arxiv.org/abs/2007.05517} {arXiv:2007.05517 [astro-ph.CO]}
  \BibitemShut {NoStop}%
\bibitem [{\citenamefont {Su}\ \emph {et~al.}(2020)\citenamefont {Su},
  \citenamefont {Wang}, \citenamefont {Wu}, \citenamefont {Yang},\ and\
  \citenamefont {Zhu}}]{Su:2020zny}%
  \BibitemOpen
  \bibfield  {author} {\bibinfo {author} {\bibfnamefont {L.}~\bibnamefont
  {Su}}, \bibinfo {author} {\bibfnamefont {W.}~\bibnamefont {Wang}}, \bibinfo
  {author} {\bibfnamefont {L.}~\bibnamefont {Wu}}, \bibinfo {author}
  {\bibfnamefont {J.~M.}\ \bibnamefont {Yang}}, \ and\ \bibinfo {author}
  {\bibfnamefont {B.}~\bibnamefont {Zhu}},\ }\href {\doibase
  10.1103/PhysRevD.102.115028} {\bibfield  {journal} {\bibinfo  {journal}
  {Phys. Rev. D}\ }\textbf {\bibinfo {volume} {102}},\ \bibinfo {pages}
  {115028} (\bibinfo {year} {2020})},\ \Eprint
  {http://arxiv.org/abs/2006.11837} {arXiv:2006.11837 [hep-ph]} \BibitemShut
  {NoStop}%
\bibitem [{\citenamefont {Cao}\ \emph {et~al.}(2021)\citenamefont {Cao},
  \citenamefont {Ding},\ and\ \citenamefont {Xiang}}]{Cao:2020bwd}%
  \BibitemOpen
  \bibfield  {author} {\bibinfo {author} {\bibfnamefont {Q.-H.}\ \bibnamefont
  {Cao}}, \bibinfo {author} {\bibfnamefont {R.}~\bibnamefont {Ding}}, \ and\
  \bibinfo {author} {\bibfnamefont {Q.-F.}\ \bibnamefont {Xiang}},\ }\href
  {\doibase 10.1088/1674-1137/abe195} {\bibfield  {journal} {\bibinfo
  {journal} {Chin. Phys. C}\ }\textbf {\bibinfo {volume} {45}},\ \bibinfo
  {pages} {045002} (\bibinfo {year} {2021})},\ \Eprint
  {http://arxiv.org/abs/2006.12767} {arXiv:2006.12767 [hep-ph]} \BibitemShut
  {NoStop}%
\bibitem [{\citenamefont {Ge}\ \emph {et~al.}(2020)\citenamefont {Ge},
  \citenamefont {Pasquini},\ and\ \citenamefont {Sheng}}]{Ge:2020jfn}%
  \BibitemOpen
  \bibfield  {author} {\bibinfo {author} {\bibfnamefont {S.-F.}\ \bibnamefont
  {Ge}}, \bibinfo {author} {\bibfnamefont {P.}~\bibnamefont {Pasquini}}, \ and\
  \bibinfo {author} {\bibfnamefont {J.}~\bibnamefont {Sheng}},\ }\href
  {\doibase 10.1016/j.physletb.2020.135787} {\bibfield  {journal} {\bibinfo
  {journal} {Phys. Lett. B}\ }\textbf {\bibinfo {volume} {810}},\ \bibinfo
  {pages} {135787} (\bibinfo {year} {2020})},\ \Eprint
  {http://arxiv.org/abs/2006.16069} {arXiv:2006.16069 [hep-ph]} \BibitemShut
  {NoStop}%
\bibitem [{\citenamefont {Choi}\ \emph {et~al.}(2021)\citenamefont {Choi},
  \citenamefont {Lee},\ and\ \citenamefont {Zhu}}]{Choi:2020ysq}%
  \BibitemOpen
  \bibfield  {author} {\bibinfo {author} {\bibfnamefont {S.-M.}\ \bibnamefont
  {Choi}}, \bibinfo {author} {\bibfnamefont {H.~M.}\ \bibnamefont {Lee}}, \
  and\ \bibinfo {author} {\bibfnamefont {B.}~\bibnamefont {Zhu}},\ }\href
  {\doibase 10.1007/JHEP04(2021)251} {\bibfield  {journal} {\bibinfo  {journal}
  {JHEP}\ }\textbf {\bibinfo {volume} {04}},\ \bibinfo {pages} {251} (\bibinfo
  {year} {2021})},\ \Eprint {http://arxiv.org/abs/2012.03713} {arXiv:2012.03713
  [hep-ph]} \BibitemShut {NoStop}%
\bibitem [{\citenamefont {An}\ \emph {et~al.}(2020)\citenamefont {An},
  \citenamefont {Pospelov}, \citenamefont {Pradler},\ and\ \citenamefont
  {Ritz}}]{An:2020bxd}%
  \BibitemOpen
  \bibfield  {author} {\bibinfo {author} {\bibfnamefont {H.}~\bibnamefont
  {An}}, \bibinfo {author} {\bibfnamefont {M.}~\bibnamefont {Pospelov}},
  \bibinfo {author} {\bibfnamefont {J.}~\bibnamefont {Pradler}}, \ and\
  \bibinfo {author} {\bibfnamefont {A.}~\bibnamefont {Ritz}},\ }\href {\doibase
  10.1103/PhysRevD.102.115022} {\bibfield  {journal} {\bibinfo  {journal}
  {Phys. Rev. D}\ }\textbf {\bibinfo {volume} {102}},\ \bibinfo {pages}
  {115022} (\bibinfo {year} {2020})},\ \Eprint
  {http://arxiv.org/abs/2006.13929} {arXiv:2006.13929 [hep-ph]} \BibitemShut
  {NoStop}%
\bibitem [{\citenamefont {Bloch}\ \emph {et~al.}(2021)\citenamefont {Bloch},
  \citenamefont {Caputo}, \citenamefont {Essig}, \citenamefont {Redigolo},
  \citenamefont {Sholapurkar},\ and\ \citenamefont {Volansky}}]{Bloch:2020uzh}%
  \BibitemOpen
  \bibfield  {author} {\bibinfo {author} {\bibfnamefont {I.~M.}\ \bibnamefont
  {Bloch}}, \bibinfo {author} {\bibfnamefont {A.}~\bibnamefont {Caputo}},
  \bibinfo {author} {\bibfnamefont {R.}~\bibnamefont {Essig}}, \bibinfo
  {author} {\bibfnamefont {D.}~\bibnamefont {Redigolo}}, \bibinfo {author}
  {\bibfnamefont {M.}~\bibnamefont {Sholapurkar}}, \ and\ \bibinfo {author}
  {\bibfnamefont {T.}~\bibnamefont {Volansky}},\ }\href {\doibase
  10.1007/JHEP01(2021)178} {\bibfield  {journal} {\bibinfo  {journal} {JHEP}\
  }\textbf {\bibinfo {volume} {01}},\ \bibinfo {pages} {178} (\bibinfo {year}
  {2021})},\ \Eprint {http://arxiv.org/abs/2006.14521} {arXiv:2006.14521
  [hep-ph]} \BibitemShut {NoStop}%
\bibitem [{\citenamefont {Zu}\ \emph {et~al.}(2021)\citenamefont {Zu},
  \citenamefont {Foot}, \citenamefont {Fan},\ and\ \citenamefont
  {Feng}}]{Zu:2020bsx}%
  \BibitemOpen
  \bibfield  {author} {\bibinfo {author} {\bibfnamefont {L.}~\bibnamefont
  {Zu}}, \bibinfo {author} {\bibfnamefont {R.}~\bibnamefont {Foot}}, \bibinfo
  {author} {\bibfnamefont {Y.-Z.}\ \bibnamefont {Fan}}, \ and\ \bibinfo
  {author} {\bibfnamefont {L.}~\bibnamefont {Feng}},\ }\href {\doibase
  10.1088/1475-7516/2021/01/070} {\bibfield  {journal} {\bibinfo  {journal}
  {JCAP}\ }\textbf {\bibinfo {volume} {01}},\ \bibinfo {pages} {070} (\bibinfo
  {year} {2021})},\ \Eprint {http://arxiv.org/abs/2007.15191} {arXiv:2007.15191
  [hep-ph]} \BibitemShut {NoStop}%
\bibitem [{\citenamefont {Guo}\ \emph {et~al.}(2020)\citenamefont {Guo},
  \citenamefont {Tsai}, \citenamefont {Wu},\ and\ \citenamefont
  {Yuan}}]{Guo:2020oum}%
  \BibitemOpen
  \bibfield  {author} {\bibinfo {author} {\bibfnamefont {G.}~\bibnamefont
  {Guo}}, \bibinfo {author} {\bibfnamefont {Y.-L.~S.}\ \bibnamefont {Tsai}},
  \bibinfo {author} {\bibfnamefont {M.-R.}\ \bibnamefont {Wu}}, \ and\ \bibinfo
  {author} {\bibfnamefont {Q.}~\bibnamefont {Yuan}},\ }\href {\doibase
  10.1103/PhysRevD.102.103004} {\bibfield  {journal} {\bibinfo  {journal}
  {Phys. Rev. D}\ }\textbf {\bibinfo {volume} {102}},\ \bibinfo {pages}
  {103004} (\bibinfo {year} {2020})},\ \Eprint
  {http://arxiv.org/abs/2008.12137} {arXiv:2008.12137 [astro-ph.HE]}
  \BibitemShut {NoStop}%
\bibitem [{\citenamefont {Du}\ \emph {et~al.}(2021)\citenamefont {Du},
  \citenamefont {Liang}, \citenamefont {Liu}, \citenamefont {Tran},\ and\
  \citenamefont {Xue}}]{Du:2020ybt}%
  \BibitemOpen
  \bibfield  {author} {\bibinfo {author} {\bibfnamefont {M.}~\bibnamefont
  {Du}}, \bibinfo {author} {\bibfnamefont {J.}~\bibnamefont {Liang}}, \bibinfo
  {author} {\bibfnamefont {Z.}~\bibnamefont {Liu}}, \bibinfo {author}
  {\bibfnamefont {V.~Q.}\ \bibnamefont {Tran}}, \ and\ \bibinfo {author}
  {\bibfnamefont {Y.}~\bibnamefont {Xue}},\ }\href {\doibase
  10.1088/1674-1137/abc244} {\bibfield  {journal} {\bibinfo  {journal} {Chin.
  Phys. C}\ }\textbf {\bibinfo {volume} {45}},\ \bibinfo {pages} {013114}
  (\bibinfo {year} {2021})},\ \Eprint {http://arxiv.org/abs/2006.11949}
  {arXiv:2006.11949 [hep-ph]} \BibitemShut {NoStop}%
\bibitem [{\citenamefont {Knapen}\ \emph
  {et~al.}(2021{\natexlab{a}})\citenamefont {Knapen}, \citenamefont
  {Kozaczuk},\ and\ \citenamefont {Lin}}]{Knapen:2021run}%
  \BibitemOpen
  \bibfield  {author} {\bibinfo {author} {\bibfnamefont {S.}~\bibnamefont
  {Knapen}}, \bibinfo {author} {\bibfnamefont {J.}~\bibnamefont {Kozaczuk}}, \
  and\ \bibinfo {author} {\bibfnamefont {T.}~\bibnamefont {Lin}},\ }\href
  {\doibase 10.1103/PhysRevD.104.015031} {\bibfield  {journal} {\bibinfo
  {journal} {Phys. Rev. D}\ }\textbf {\bibinfo {volume} {104}},\ \bibinfo
  {pages} {015031} (\bibinfo {year} {2021}{\natexlab{a}})},\ \Eprint
  {http://arxiv.org/abs/2101.08275} {arXiv:2101.08275 [hep-ph]} \BibitemShut
  {NoStop}%
\bibitem [{\citenamefont {Chao}\ \emph {et~al.}(2021)\citenamefont {Chao},
  \citenamefont {Jin},\ and\ \citenamefont {Peng}}]{Chao:2021liw}%
  \BibitemOpen
  \bibfield  {author} {\bibinfo {author} {\bibfnamefont {W.}~\bibnamefont
  {Chao}}, \bibinfo {author} {\bibfnamefont {M.}~\bibnamefont {Jin}}, \ and\
  \bibinfo {author} {\bibfnamefont {Y.-Q.}\ \bibnamefont {Peng}},\ }\href@noop
  {} {\  (\bibinfo {year} {2021})},\ \Eprint {http://arxiv.org/abs/2109.14944}
  {arXiv:2109.14944 [hep-ph]} \BibitemShut {NoStop}%
\bibitem [{\citenamefont {Hochberg}\ \emph {et~al.}(2021)\citenamefont
  {Hochberg}, \citenamefont {Kahn}, \citenamefont {Kurinsky}, \citenamefont
  {Lehmann}, \citenamefont {Yu},\ and\ \citenamefont
  {Berggren}}]{Hochberg:2021pkt}%
  \BibitemOpen
  \bibfield  {author} {\bibinfo {author} {\bibfnamefont {Y.}~\bibnamefont
  {Hochberg}}, \bibinfo {author} {\bibfnamefont {Y.}~\bibnamefont {Kahn}},
  \bibinfo {author} {\bibfnamefont {N.}~\bibnamefont {Kurinsky}}, \bibinfo
  {author} {\bibfnamefont {B.~V.}\ \bibnamefont {Lehmann}}, \bibinfo {author}
  {\bibfnamefont {T.~C.}\ \bibnamefont {Yu}}, \ and\ \bibinfo {author}
  {\bibfnamefont {K.~K.}\ \bibnamefont {Berggren}},\ }\href {\doibase
  10.1103/PhysRevLett.127.151802} {\bibfield  {journal} {\bibinfo  {journal}
  {Phys. Rev. Lett.}\ }\textbf {\bibinfo {volume} {127}},\ \bibinfo {pages}
  {151802} (\bibinfo {year} {2021})},\ \Eprint
  {http://arxiv.org/abs/2101.08263} {arXiv:2101.08263 [hep-ph]} \BibitemShut
  {NoStop}%
\bibitem [{\citenamefont {Migdal}(1939)}]{Migdal:1939}%
  \BibitemOpen
  \bibfield  {author} {\bibinfo {author} {\bibfnamefont {A.}~\bibnamefont
  {Migdal}},\ }\href@noop {} {\bibfield  {journal} {\bibinfo  {journal}
  {Sov.Phys.JETP}\ }\textbf {\bibinfo {volume} {9}},\ \bibinfo {pages} {1163}
  (\bibinfo {year} {1939})}\BibitemShut {NoStop}%
\bibitem [{\citenamefont {Vergados}\ and\ \citenamefont
  {Ejiri}(2005)}]{Vergados:2005dpd}%
  \BibitemOpen
  \bibfield  {author} {\bibinfo {author} {\bibfnamefont {J.~D.}\ \bibnamefont
  {Vergados}}\ and\ \bibinfo {author} {\bibfnamefont {H.}~\bibnamefont
  {Ejiri}},\ }\href {\doibase 10.1016/j.physletb.2004.11.085} {\bibfield
  {journal} {\bibinfo  {journal} {Phys. Lett. B}\ }\textbf {\bibinfo {volume}
  {606}},\ \bibinfo {pages} {313} (\bibinfo {year} {2005})},\ \Eprint
  {http://arxiv.org/abs/hep-ph/0401151} {arXiv:hep-ph/0401151} \BibitemShut
  {NoStop}%
\bibitem [{\citenamefont {Moustakidis}\ \emph {et~al.}(2005)\citenamefont
  {Moustakidis}, \citenamefont {Vergados},\ and\ \citenamefont
  {Ejiri}}]{Moustakidis:2005gx}%
  \BibitemOpen
  \bibfield  {author} {\bibinfo {author} {\bibfnamefont {C.~C.}\ \bibnamefont
  {Moustakidis}}, \bibinfo {author} {\bibfnamefont {J.~D.}\ \bibnamefont
  {Vergados}}, \ and\ \bibinfo {author} {\bibfnamefont {H.}~\bibnamefont
  {Ejiri}},\ }\href {\doibase 10.1016/j.nuclphysb.2005.08.033} {\bibfield
  {journal} {\bibinfo  {journal} {Nucl. Phys. B}\ }\textbf {\bibinfo {volume}
  {727}},\ \bibinfo {pages} {406} (\bibinfo {year} {2005})},\ \Eprint
  {http://arxiv.org/abs/hep-ph/0507123} {arXiv:hep-ph/0507123} \BibitemShut
  {NoStop}%
\bibitem [{\citenamefont {Ejiri}\ \emph {et~al.}(2006)\citenamefont {Ejiri},
  \citenamefont {Moustakidis},\ and\ \citenamefont {Vergados}}]{Ejiri:2005aj}%
  \BibitemOpen
  \bibfield  {author} {\bibinfo {author} {\bibfnamefont {H.}~\bibnamefont
  {Ejiri}}, \bibinfo {author} {\bibfnamefont {C.~C.}\ \bibnamefont
  {Moustakidis}}, \ and\ \bibinfo {author} {\bibfnamefont {J.~D.}\ \bibnamefont
  {Vergados}},\ }\href {\doibase 10.1016/j.physletb.2006.03.037} {\bibfield
  {journal} {\bibinfo  {journal} {Phys. Lett. B}\ }\textbf {\bibinfo {volume}
  {639}},\ \bibinfo {pages} {218} (\bibinfo {year} {2006})},\ \Eprint
  {http://arxiv.org/abs/hep-ph/0510042} {arXiv:hep-ph/0510042} \BibitemShut
  {NoStop}%
\bibitem [{\citenamefont {Bernabei}\ \emph {et~al.}(2007)\citenamefont
  {Bernabei} \emph {et~al.}}]{Bernabei:2007jz}%
  \BibitemOpen
  \bibfield  {author} {\bibinfo {author} {\bibfnamefont {R.}~\bibnamefont
  {Bernabei}} \emph {et~al.},\ }\href {\doibase 10.1142/S0217751X07037093}
  {\bibfield  {journal} {\bibinfo  {journal} {Int. J. Mod. Phys. A}\ }\textbf
  {\bibinfo {volume} {22}},\ \bibinfo {pages} {3155} (\bibinfo {year}
  {2007})},\ \Eprint {http://arxiv.org/abs/0706.1421} {arXiv:0706.1421
  [astro-ph]} \BibitemShut {NoStop}%
\bibitem [{\citenamefont {Ibe}\ \emph {et~al.}(2018)\citenamefont {Ibe},
  \citenamefont {Nakano}, \citenamefont {Shoji},\ and\ \citenamefont
  {Suzuki}}]{Ibe:2017yqa}%
  \BibitemOpen
  \bibfield  {author} {\bibinfo {author} {\bibfnamefont {M.}~\bibnamefont
  {Ibe}}, \bibinfo {author} {\bibfnamefont {W.}~\bibnamefont {Nakano}},
  \bibinfo {author} {\bibfnamefont {Y.}~\bibnamefont {Shoji}}, \ and\ \bibinfo
  {author} {\bibfnamefont {K.}~\bibnamefont {Suzuki}},\ }\href {\doibase
  10.1007/JHEP03(2018)194} {\bibfield  {journal} {\bibinfo  {journal} {JHEP}\
  }\textbf {\bibinfo {volume} {03}},\ \bibinfo {pages} {194} (\bibinfo {year}
  {2018})},\ \Eprint {http://arxiv.org/abs/1707.07258} {arXiv:1707.07258
  [hep-ph]} \BibitemShut {NoStop}%
\bibitem [{\citenamefont {Dolan}\ \emph {et~al.}(2018)\citenamefont {Dolan},
  \citenamefont {Kahlhoefer},\ and\ \citenamefont {McCabe}}]{Dolan:2017xbu}%
  \BibitemOpen
  \bibfield  {author} {\bibinfo {author} {\bibfnamefont {M.~J.}\ \bibnamefont
  {Dolan}}, \bibinfo {author} {\bibfnamefont {F.}~\bibnamefont {Kahlhoefer}}, \
  and\ \bibinfo {author} {\bibfnamefont {C.}~\bibnamefont {McCabe}},\ }\href
  {\doibase 10.1103/PhysRevLett.121.101801} {\bibfield  {journal} {\bibinfo
  {journal} {Phys. Rev. Lett.}\ }\textbf {\bibinfo {volume} {121}},\ \bibinfo
  {pages} {101801} (\bibinfo {year} {2018})},\ \Eprint
  {http://arxiv.org/abs/1711.09906} {arXiv:1711.09906 [hep-ph]} \BibitemShut
  {NoStop}%
\bibitem [{\citenamefont {Bell}\ \emph {et~al.}(2020)\citenamefont {Bell},
  \citenamefont {Dent}, \citenamefont {Newstead}, \citenamefont {Sabharwal},\
  and\ \citenamefont {Weiler}}]{Bell:2019egg}%
  \BibitemOpen
  \bibfield  {author} {\bibinfo {author} {\bibfnamefont {N.~F.}\ \bibnamefont
  {Bell}}, \bibinfo {author} {\bibfnamefont {J.~B.}\ \bibnamefont {Dent}},
  \bibinfo {author} {\bibfnamefont {J.~L.}\ \bibnamefont {Newstead}}, \bibinfo
  {author} {\bibfnamefont {S.}~\bibnamefont {Sabharwal}}, \ and\ \bibinfo
  {author} {\bibfnamefont {T.~J.}\ \bibnamefont {Weiler}},\ }\href {\doibase
  10.1103/PhysRevD.101.015012} {\bibfield  {journal} {\bibinfo  {journal}
  {Phys. Rev. D}\ }\textbf {\bibinfo {volume} {101}},\ \bibinfo {pages}
  {015012} (\bibinfo {year} {2020})},\ \Eprint
  {http://arxiv.org/abs/1905.00046} {arXiv:1905.00046 [hep-ph]} \BibitemShut
  {NoStop}%
\bibitem [{\citenamefont {Essig}\ \emph {et~al.}(2020)\citenamefont {Essig},
  \citenamefont {Pradler}, \citenamefont {Sholapurkar},\ and\ \citenamefont
  {Yu}}]{Essig:2019xkx}%
  \BibitemOpen
  \bibfield  {author} {\bibinfo {author} {\bibfnamefont {R.}~\bibnamefont
  {Essig}}, \bibinfo {author} {\bibfnamefont {J.}~\bibnamefont {Pradler}},
  \bibinfo {author} {\bibfnamefont {M.}~\bibnamefont {Sholapurkar}}, \ and\
  \bibinfo {author} {\bibfnamefont {T.-T.}\ \bibnamefont {Yu}},\ }\href
  {\doibase 10.1103/PhysRevLett.124.021801} {\bibfield  {journal} {\bibinfo
  {journal} {Phys. Rev. Lett.}\ }\textbf {\bibinfo {volume} {124}},\ \bibinfo
  {pages} {021801} (\bibinfo {year} {2020})},\ \Eprint
  {http://arxiv.org/abs/1908.10881} {arXiv:1908.10881 [hep-ph]} \BibitemShut
  {NoStop}%
\bibitem [{\citenamefont {Baxter}\ \emph {et~al.}(2020)\citenamefont {Baxter},
  \citenamefont {Kahn},\ and\ \citenamefont {Krnjaic}}]{Baxter:2019pnz}%
  \BibitemOpen
  \bibfield  {author} {\bibinfo {author} {\bibfnamefont {D.}~\bibnamefont
  {Baxter}}, \bibinfo {author} {\bibfnamefont {Y.}~\bibnamefont {Kahn}}, \ and\
  \bibinfo {author} {\bibfnamefont {G.}~\bibnamefont {Krnjaic}},\ }\href
  {\doibase 10.1103/PhysRevD.101.076014} {\bibfield  {journal} {\bibinfo
  {journal} {Phys. Rev. D}\ }\textbf {\bibinfo {volume} {101}},\ \bibinfo
  {pages} {076014} (\bibinfo {year} {2020})},\ \Eprint
  {http://arxiv.org/abs/1908.00012} {arXiv:1908.00012 [hep-ph]} \BibitemShut
  {NoStop}%
\bibitem [{\citenamefont {Grilli~di Cortona}\ \emph {et~al.}(2020)\citenamefont
  {Grilli~di Cortona}, \citenamefont {Messina},\ and\ \citenamefont
  {Piacentini}}]{GrillidiCortona:2020owp}%
  \BibitemOpen
  \bibfield  {author} {\bibinfo {author} {\bibfnamefont {G.}~\bibnamefont
  {Grilli~di Cortona}}, \bibinfo {author} {\bibfnamefont {A.}~\bibnamefont
  {Messina}}, \ and\ \bibinfo {author} {\bibfnamefont {S.}~\bibnamefont
  {Piacentini}},\ }\href {\doibase 10.1007/JHEP11(2020)034} {\bibfield
  {journal} {\bibinfo  {journal} {JHEP}\ }\textbf {\bibinfo {volume} {11}},\
  \bibinfo {pages} {034} (\bibinfo {year} {2020})},\ \Eprint
  {http://arxiv.org/abs/2006.02453} {arXiv:2006.02453 [hep-ph]} \BibitemShut
  {NoStop}%
\bibitem [{\citenamefont {Liu}\ \emph {et~al.}(2020)\citenamefont {Liu},
  \citenamefont {Wu}, \citenamefont {Chi},\ and\ \citenamefont
  {Chen}}]{Liu:2020pat}%
  \BibitemOpen
  \bibfield  {author} {\bibinfo {author} {\bibfnamefont {C.~P.}\ \bibnamefont
  {Liu}}, \bibinfo {author} {\bibfnamefont {C.-P.}\ \bibnamefont {Wu}},
  \bibinfo {author} {\bibfnamefont {H.-C.}\ \bibnamefont {Chi}}, \ and\
  \bibinfo {author} {\bibfnamefont {J.-W.}\ \bibnamefont {Chen}},\ }\href
  {\doibase 10.1103/PhysRevD.102.121303} {\bibfield  {journal} {\bibinfo
  {journal} {Phys. Rev. D}\ }\textbf {\bibinfo {volume} {102}},\ \bibinfo
  {pages} {121303} (\bibinfo {year} {2020})},\ \Eprint
  {http://arxiv.org/abs/2007.10965} {arXiv:2007.10965 [hep-ph]} \BibitemShut
  {NoStop}%
\bibitem [{\citenamefont {Knapen}\ \emph
  {et~al.}(2021{\natexlab{b}})\citenamefont {Knapen}, \citenamefont
  {Kozaczuk},\ and\ \citenamefont {Lin}}]{Knapen:2020aky}%
  \BibitemOpen
  \bibfield  {author} {\bibinfo {author} {\bibfnamefont {S.}~\bibnamefont
  {Knapen}}, \bibinfo {author} {\bibfnamefont {J.}~\bibnamefont {Kozaczuk}}, \
  and\ \bibinfo {author} {\bibfnamefont {T.}~\bibnamefont {Lin}},\ }\href
  {\doibase 10.1103/PhysRevLett.127.081805} {\bibfield  {journal} {\bibinfo
  {journal} {Phys. Rev. Lett.}\ }\textbf {\bibinfo {volume} {127}},\ \bibinfo
  {pages} {081805} (\bibinfo {year} {2021}{\natexlab{b}})},\ \Eprint
  {http://arxiv.org/abs/2011.09496} {arXiv:2011.09496 [hep-ph]} \BibitemShut
  {NoStop}%
\bibitem [{\citenamefont {Flambaum}\ \emph {et~al.}(2020)\citenamefont
  {Flambaum}, \citenamefont {Su}, \citenamefont {Wu},\ and\ \citenamefont
  {Zhu}}]{Flambaum:2020xxo}%
  \BibitemOpen
  \bibfield  {author} {\bibinfo {author} {\bibfnamefont {V.~V.}\ \bibnamefont
  {Flambaum}}, \bibinfo {author} {\bibfnamefont {L.}~\bibnamefont {Su}},
  \bibinfo {author} {\bibfnamefont {L.}~\bibnamefont {Wu}}, \ and\ \bibinfo
  {author} {\bibfnamefont {B.}~\bibnamefont {Zhu}},\ }\href@noop {} {\
  (\bibinfo {year} {2020})},\ \Eprint {http://arxiv.org/abs/2012.09751}
  {arXiv:2012.09751 [hep-ph]} \BibitemShut {NoStop}%
\bibitem [{\citenamefont {He}\ \emph {et~al.}(2020)\citenamefont {He},
  \citenamefont {Wang},\ and\ \citenamefont {Zheng}}]{He:2020sat}%
  \BibitemOpen
  \bibfield  {author} {\bibinfo {author} {\bibfnamefont {H.-J.}\ \bibnamefont
  {He}}, \bibinfo {author} {\bibfnamefont {Y.-C.}\ \bibnamefont {Wang}}, \ and\
  \bibinfo {author} {\bibfnamefont {J.}~\bibnamefont {Zheng}},\ }\href@noop {}
  {\  (\bibinfo {year} {2020})},\ \Eprint {http://arxiv.org/abs/2012.05891}
  {arXiv:2012.05891 [hep-ph]} \BibitemShut {NoStop}%
\bibitem [{\citenamefont {Liang}\ \emph {et~al.}(2021)\citenamefont {Liang},
  \citenamefont {Mo}, \citenamefont {Zheng},\ and\ \citenamefont
  {Zhang}}]{Liang:2020ryg}%
  \BibitemOpen
  \bibfield  {author} {\bibinfo {author} {\bibfnamefont {Z.-L.}\ \bibnamefont
  {Liang}}, \bibinfo {author} {\bibfnamefont {C.}~\bibnamefont {Mo}}, \bibinfo
  {author} {\bibfnamefont {F.}~\bibnamefont {Zheng}}, \ and\ \bibinfo {author}
  {\bibfnamefont {P.}~\bibnamefont {Zhang}},\ }\href {\doibase
  10.1103/PhysRevD.104.056009} {\bibfield  {journal} {\bibinfo  {journal}
  {Phys. Rev. D}\ }\textbf {\bibinfo {volume} {104}},\ \bibinfo {pages}
  {056009} (\bibinfo {year} {2021})},\ \Eprint
  {http://arxiv.org/abs/2011.13352} {arXiv:2011.13352 [hep-ph]} \BibitemShut
  {NoStop}%
\bibitem [{\citenamefont {Bell}\ \emph {et~al.}(2021)\citenamefont {Bell},
  \citenamefont {Dent}, \citenamefont {Dutta}, \citenamefont {Ghosh},
  \citenamefont {Kumar},\ and\ \citenamefont {Newstead}}]{Bell:2021zkr}%
  \BibitemOpen
  \bibfield  {author} {\bibinfo {author} {\bibfnamefont {N.~F.}\ \bibnamefont
  {Bell}}, \bibinfo {author} {\bibfnamefont {J.~B.}\ \bibnamefont {Dent}},
  \bibinfo {author} {\bibfnamefont {B.}~\bibnamefont {Dutta}}, \bibinfo
  {author} {\bibfnamefont {S.}~\bibnamefont {Ghosh}}, \bibinfo {author}
  {\bibfnamefont {J.}~\bibnamefont {Kumar}}, \ and\ \bibinfo {author}
  {\bibfnamefont {J.~L.}\ \bibnamefont {Newstead}},\ }\href {\doibase
  10.1103/PhysRevD.104.076013} {\bibfield  {journal} {\bibinfo  {journal}
  {Phys. Rev. D}\ }\textbf {\bibinfo {volume} {104}},\ \bibinfo {pages}
  {076013} (\bibinfo {year} {2021})},\ \Eprint
  {http://arxiv.org/abs/2103.05890} {arXiv:2103.05890 [hep-ph]} \BibitemShut
  {NoStop}%
\bibitem [{\citenamefont {Acevedo}\ \emph {et~al.}(2021)\citenamefont
  {Acevedo}, \citenamefont {Bramante},\ and\ \citenamefont
  {Goodman}}]{Acevedo:2021kly}%
  \BibitemOpen
  \bibfield  {author} {\bibinfo {author} {\bibfnamefont {J.~F.}\ \bibnamefont
  {Acevedo}}, \bibinfo {author} {\bibfnamefont {J.}~\bibnamefont {Bramante}}, \
  and\ \bibinfo {author} {\bibfnamefont {A.}~\bibnamefont {Goodman}},\
  }\href@noop {} {\  (\bibinfo {year} {2021})},\ \Eprint
  {http://arxiv.org/abs/2108.10889} {arXiv:2108.10889 [hep-ph]} \BibitemShut
  {NoStop}%
\bibitem [{\citenamefont {Liu}\ \emph {et~al.}(2019)\citenamefont {Liu} \emph
  {et~al.}}]{CDEX:2019hzn}%
  \BibitemOpen
  \bibfield  {author} {\bibinfo {author} {\bibfnamefont {Z.~Z.}\ \bibnamefont
  {Liu}} \emph {et~al.} (\bibinfo {collaboration} {CDEX}),\ }\href {\doibase
  10.1103/PhysRevLett.123.161301} {\bibfield  {journal} {\bibinfo  {journal}
  {Phys. Rev. Lett.}\ }\textbf {\bibinfo {volume} {123}},\ \bibinfo {pages}
  {161301} (\bibinfo {year} {2019})},\ \Eprint
  {http://arxiv.org/abs/1905.00354} {arXiv:1905.00354 [hep-ex]} \BibitemShut
  {NoStop}%
\bibitem [{\citenamefont {Aprile}\ \emph
  {et~al.}(2019{\natexlab{a}})\citenamefont {Aprile} \emph
  {et~al.}}]{XENON:2019zpr}%
  \BibitemOpen
  \bibfield  {author} {\bibinfo {author} {\bibfnamefont {E.}~\bibnamefont
  {Aprile}} \emph {et~al.} (\bibinfo {collaboration} {XENON}),\ }\href
  {\doibase 10.1103/PhysRevLett.123.241803} {\bibfield  {journal} {\bibinfo
  {journal} {Phys. Rev. Lett.}\ }\textbf {\bibinfo {volume} {123}},\ \bibinfo
  {pages} {241803} (\bibinfo {year} {2019}{\natexlab{a}})},\ \Eprint
  {http://arxiv.org/abs/1907.12771} {arXiv:1907.12771 [hep-ex]} \BibitemShut
  {NoStop}%
\bibitem [{\citenamefont {Adhikari}\ \emph {et~al.}(2021)\citenamefont
  {Adhikari} \emph {et~al.}}]{COSINE-100:2021poy}%
  \BibitemOpen
  \bibfield  {author} {\bibinfo {author} {\bibfnamefont {G.}~\bibnamefont
  {Adhikari}} \emph {et~al.} (\bibinfo {collaboration} {COSINE-100}),\
  }\href@noop {} {\  (\bibinfo {year} {2021})},\ \Eprint
  {http://arxiv.org/abs/2110.05806} {arXiv:2110.05806 [hep-ex]} \BibitemShut
  {NoStop}%
\bibitem [{\citenamefont {Agrawal}\ \emph {et~al.}(2010)\citenamefont
  {Agrawal}, \citenamefont {Chacko}, \citenamefont {Kilic},\ and\ \citenamefont
  {Mishra}}]{Agrawal:2010fh}%
  \BibitemOpen
  \bibfield  {author} {\bibinfo {author} {\bibfnamefont {P.}~\bibnamefont
  {Agrawal}}, \bibinfo {author} {\bibfnamefont {Z.}~\bibnamefont {Chacko}},
  \bibinfo {author} {\bibfnamefont {C.}~\bibnamefont {Kilic}}, \ and\ \bibinfo
  {author} {\bibfnamefont {R.~K.}\ \bibnamefont {Mishra}},\ }\href@noop {} {\
  (\bibinfo {year} {2010})},\ \Eprint {http://arxiv.org/abs/1003.1912}
  {arXiv:1003.1912 [hep-ph]} \BibitemShut {NoStop}%
\bibitem [{\citenamefont {Fan}\ \emph {et~al.}(2010)\citenamefont {Fan},
  \citenamefont {Reece},\ and\ \citenamefont {Wang}}]{Fan:2010gt}%
  \BibitemOpen
  \bibfield  {author} {\bibinfo {author} {\bibfnamefont {J.}~\bibnamefont
  {Fan}}, \bibinfo {author} {\bibfnamefont {M.}~\bibnamefont {Reece}}, \ and\
  \bibinfo {author} {\bibfnamefont {L.-T.}\ \bibnamefont {Wang}},\ }\href
  {\doibase 10.1088/1475-7516/2010/11/042} {\bibfield  {journal} {\bibinfo
  {journal} {JCAP}\ }\textbf {\bibinfo {volume} {11}},\ \bibinfo {pages} {042}
  (\bibinfo {year} {2010})},\ \Eprint {http://arxiv.org/abs/1008.1591}
  {arXiv:1008.1591 [hep-ph]} \BibitemShut {NoStop}%
\bibitem [{\citenamefont {Freytsis}\ and\ \citenamefont
  {Ligeti}(2011)}]{Freytsis:2010ne}%
  \BibitemOpen
  \bibfield  {author} {\bibinfo {author} {\bibfnamefont {M.}~\bibnamefont
  {Freytsis}}\ and\ \bibinfo {author} {\bibfnamefont {Z.}~\bibnamefont
  {Ligeti}},\ }\href {\doibase 10.1103/PhysRevD.83.115009} {\bibfield
  {journal} {\bibinfo  {journal} {Phys. Rev. D}\ }\textbf {\bibinfo {volume}
  {83}},\ \bibinfo {pages} {115009} (\bibinfo {year} {2011})},\ \Eprint
  {http://arxiv.org/abs/1012.5317} {arXiv:1012.5317 [hep-ph]} \BibitemShut
  {NoStop}%
\bibitem [{\citenamefont {Fitzpatrick}\ \emph {et~al.}(2013)\citenamefont
  {Fitzpatrick}, \citenamefont {Haxton}, \citenamefont {Katz}, \citenamefont
  {Lubbers},\ and\ \citenamefont {Xu}}]{Fitzpatrick:2012ix}%
  \BibitemOpen
  \bibfield  {author} {\bibinfo {author} {\bibfnamefont {A.~L.}\ \bibnamefont
  {Fitzpatrick}}, \bibinfo {author} {\bibfnamefont {W.}~\bibnamefont {Haxton}},
  \bibinfo {author} {\bibfnamefont {E.}~\bibnamefont {Katz}}, \bibinfo {author}
  {\bibfnamefont {N.}~\bibnamefont {Lubbers}}, \ and\ \bibinfo {author}
  {\bibfnamefont {Y.}~\bibnamefont {Xu}},\ }\href {\doibase
  10.1088/1475-7516/2013/02/004} {\bibfield  {journal} {\bibinfo  {journal}
  {JCAP}\ }\textbf {\bibinfo {volume} {02}},\ \bibinfo {pages} {004} (\bibinfo
  {year} {2013})},\ \Eprint {http://arxiv.org/abs/1203.3542} {arXiv:1203.3542
  [hep-ph]} \BibitemShut {NoStop}%
\bibitem [{\citenamefont {Fitzpatrick}\ \emph {et~al.}(2012)\citenamefont
  {Fitzpatrick}, \citenamefont {Haxton}, \citenamefont {Katz}, \citenamefont
  {Lubbers},\ and\ \citenamefont {Xu}}]{Fitzpatrick:2012ib}%
  \BibitemOpen
  \bibfield  {author} {\bibinfo {author} {\bibfnamefont {A.~L.}\ \bibnamefont
  {Fitzpatrick}}, \bibinfo {author} {\bibfnamefont {W.}~\bibnamefont {Haxton}},
  \bibinfo {author} {\bibfnamefont {E.}~\bibnamefont {Katz}}, \bibinfo {author}
  {\bibfnamefont {N.}~\bibnamefont {Lubbers}}, \ and\ \bibinfo {author}
  {\bibfnamefont {Y.}~\bibnamefont {Xu}},\ }\href@noop {} {\  (\bibinfo {year}
  {2012})},\ \Eprint {http://arxiv.org/abs/1211.2818} {arXiv:1211.2818
  [hep-ph]} \BibitemShut {NoStop}%
\bibitem [{\citenamefont {Anand}\ \emph {et~al.}(2014)\citenamefont {Anand},
  \citenamefont {Fitzpatrick},\ and\ \citenamefont {Haxton}}]{Anand:2013yka}%
  \BibitemOpen
  \bibfield  {author} {\bibinfo {author} {\bibfnamefont {N.}~\bibnamefont
  {Anand}}, \bibinfo {author} {\bibfnamefont {A.~L.}\ \bibnamefont
  {Fitzpatrick}}, \ and\ \bibinfo {author} {\bibfnamefont {W.~C.}\ \bibnamefont
  {Haxton}},\ }\href {\doibase 10.1103/PhysRevC.89.065501} {\bibfield
  {journal} {\bibinfo  {journal} {Phys. Rev. C}\ }\textbf {\bibinfo {volume}
  {89}},\ \bibinfo {pages} {065501} (\bibinfo {year} {2014})},\ \Eprint
  {http://arxiv.org/abs/1308.6288} {arXiv:1308.6288 [hep-ph]} \BibitemShut
  {NoStop}%
\bibitem [{\citenamefont {Cheung}\ \emph {et~al.}(2013)\citenamefont {Cheung},
  \citenamefont {Hall}, \citenamefont {Pinner},\ and\ \citenamefont
  {Ruderman}}]{Cheung:2012qy}%
  \BibitemOpen
  \bibfield  {author} {\bibinfo {author} {\bibfnamefont {C.}~\bibnamefont
  {Cheung}}, \bibinfo {author} {\bibfnamefont {L.~J.}\ \bibnamefont {Hall}},
  \bibinfo {author} {\bibfnamefont {D.}~\bibnamefont {Pinner}}, \ and\ \bibinfo
  {author} {\bibfnamefont {J.~T.}\ \bibnamefont {Ruderman}},\ }\href {\doibase
  10.1007/JHEP05(2013)100} {\bibfield  {journal} {\bibinfo  {journal} {JHEP}\
  }\textbf {\bibinfo {volume} {05}},\ \bibinfo {pages} {100} (\bibinfo {year}
  {2013})},\ \Eprint {http://arxiv.org/abs/1211.4873} {arXiv:1211.4873
  [hep-ph]} \BibitemShut {NoStop}%
\bibitem [{\citenamefont {Huang}\ and\ \citenamefont
  {Wagner}(2014)}]{Huang:2014xua}%
  \BibitemOpen
  \bibfield  {author} {\bibinfo {author} {\bibfnamefont {P.}~\bibnamefont
  {Huang}}\ and\ \bibinfo {author} {\bibfnamefont {C.~E.~M.}\ \bibnamefont
  {Wagner}},\ }\href {\doibase 10.1103/PhysRevD.90.015018} {\bibfield
  {journal} {\bibinfo  {journal} {Phys. Rev. D}\ }\textbf {\bibinfo {volume}
  {90}},\ \bibinfo {pages} {015018} (\bibinfo {year} {2014})},\ \Eprint
  {http://arxiv.org/abs/1404.0392} {arXiv:1404.0392 [hep-ph]} \BibitemShut
  {NoStop}%
\bibitem [{\citenamefont {Han}\ \emph {et~al.}(2017)\citenamefont {Han},
  \citenamefont {Kling}, \citenamefont {Su},\ and\ \citenamefont
  {Wu}}]{Han:2016qtc}%
  \BibitemOpen
  \bibfield  {author} {\bibinfo {author} {\bibfnamefont {T.}~\bibnamefont
  {Han}}, \bibinfo {author} {\bibfnamefont {F.}~\bibnamefont {Kling}}, \bibinfo
  {author} {\bibfnamefont {S.}~\bibnamefont {Su}}, \ and\ \bibinfo {author}
  {\bibfnamefont {Y.}~\bibnamefont {Wu}},\ }\href {\doibase
  10.1007/JHEP02(2017)057} {\bibfield  {journal} {\bibinfo  {journal} {JHEP}\
  }\textbf {\bibinfo {volume} {02}},\ \bibinfo {pages} {057} (\bibinfo {year}
  {2017})},\ \Eprint {http://arxiv.org/abs/1612.02387} {arXiv:1612.02387
  [hep-ph]} \BibitemShut {NoStop}%
\bibitem [{\citenamefont {Abdughani}\ \emph {et~al.}(2018)\citenamefont
  {Abdughani}, \citenamefont {Wu},\ and\ \citenamefont
  {Yang}}]{Abdughani:2017dqs}%
  \BibitemOpen
  \bibfield  {author} {\bibinfo {author} {\bibfnamefont {M.}~\bibnamefont
  {Abdughani}}, \bibinfo {author} {\bibfnamefont {L.}~\bibnamefont {Wu}}, \
  and\ \bibinfo {author} {\bibfnamefont {J.~M.}\ \bibnamefont {Yang}},\ }\href
  {\doibase 10.1140/epjc/s10052-017-5485-2} {\bibfield  {journal} {\bibinfo
  {journal} {Eur. Phys. J. C}\ }\textbf {\bibinfo {volume} {78}},\ \bibinfo
  {pages} {4} (\bibinfo {year} {2018})},\ \Eprint
  {http://arxiv.org/abs/1705.09164} {arXiv:1705.09164 [hep-ph]} \BibitemShut
  {NoStop}%
\bibitem [{\citenamefont {Li}\ \emph {et~al.}(2015)\citenamefont {Li},
  \citenamefont {Miao},\ and\ \citenamefont {Zhou}}]{Li:2014vza}%
  \BibitemOpen
  \bibfield  {author} {\bibinfo {author} {\bibfnamefont {T.}~\bibnamefont
  {Li}}, \bibinfo {author} {\bibfnamefont {S.}~\bibnamefont {Miao}}, \ and\
  \bibinfo {author} {\bibfnamefont {Y.-F.}\ \bibnamefont {Zhou}},\ }\href
  {\doibase 10.1088/1475-7516/2015/03/032} {\bibfield  {journal} {\bibinfo
  {journal} {JCAP}\ }\textbf {\bibinfo {volume} {03}},\ \bibinfo {pages} {032}
  (\bibinfo {year} {2015})},\ \Eprint {http://arxiv.org/abs/1412.6220}
  {arXiv:1412.6220 [hep-ph]} \BibitemShut {NoStop}%
\bibitem [{\citenamefont {Ramani}\ and\ \citenamefont
  {Woolley}(2019)}]{Ramani:2019jam}%
  \BibitemOpen
  \bibfield  {author} {\bibinfo {author} {\bibfnamefont {H.}~\bibnamefont
  {Ramani}}\ and\ \bibinfo {author} {\bibfnamefont {G.}~\bibnamefont
  {Woolley}},\ }\href@noop {} {\  (\bibinfo {year} {2019})},\ \Eprint
  {http://arxiv.org/abs/1905.04319} {arXiv:1905.04319 [hep-ph]} \BibitemShut
  {NoStop}%
\bibitem [{\citenamefont {Engel}\ \emph {et~al.}(1992)\citenamefont {Engel},
  \citenamefont {Pittel},\ and\ \citenamefont {Vogel}}]{Engel:1992bf}%
  \BibitemOpen
  \bibfield  {author} {\bibinfo {author} {\bibfnamefont {J.}~\bibnamefont
  {Engel}}, \bibinfo {author} {\bibfnamefont {S.}~\bibnamefont {Pittel}}, \
  and\ \bibinfo {author} {\bibfnamefont {P.}~\bibnamefont {Vogel}},\ }\href
  {\doibase 10.1142/S0218301392000023} {\bibfield  {journal} {\bibinfo
  {journal} {Int. J. Mod. Phys. E}\ }\textbf {\bibinfo {volume} {1}},\ \bibinfo
  {pages} {1} (\bibinfo {year} {1992})}\BibitemShut {NoStop}%
\bibitem [{\citenamefont {Jungman}\ \emph {et~al.}(1996)\citenamefont
  {Jungman}, \citenamefont {Kamionkowski},\ and\ \citenamefont
  {Griest}}]{Jungman:1995df}%
  \BibitemOpen
  \bibfield  {author} {\bibinfo {author} {\bibfnamefont {G.}~\bibnamefont
  {Jungman}}, \bibinfo {author} {\bibfnamefont {M.}~\bibnamefont
  {Kamionkowski}}, \ and\ \bibinfo {author} {\bibfnamefont {K.}~\bibnamefont
  {Griest}},\ }\href {\doibase 10.1016/0370-1573(95)00058-5} {\bibfield
  {journal} {\bibinfo  {journal} {Phys. Rept.}\ }\textbf {\bibinfo {volume}
  {267}},\ \bibinfo {pages} {195} (\bibinfo {year} {1996})},\ \Eprint
  {http://arxiv.org/abs/hep-ph/9506380} {arXiv:hep-ph/9506380} \BibitemShut
  {NoStop}%
\bibitem [{\citenamefont {Hu}\ \emph {et~al.}(2021)\citenamefont {Hu},
  \citenamefont {Padua-Arg\"uelles}, \citenamefont {Leutheusser}, \citenamefont
  {Miyagi}, \citenamefont {Stroberg},\ and\ \citenamefont {Holt}}]{Hu:2021awl}%
  \BibitemOpen
  \bibfield  {author} {\bibinfo {author} {\bibfnamefont {B.~S.}\ \bibnamefont
  {Hu}}, \bibinfo {author} {\bibfnamefont {J.}~\bibnamefont
  {Padua-Arg\"uelles}}, \bibinfo {author} {\bibfnamefont {S.}~\bibnamefont
  {Leutheusser}}, \bibinfo {author} {\bibfnamefont {T.}~\bibnamefont {Miyagi}},
  \bibinfo {author} {\bibfnamefont {S.~R.}\ \bibnamefont {Stroberg}}, \ and\
  \bibinfo {author} {\bibfnamefont {J.~D.}\ \bibnamefont {Holt}},\ }\href@noop
  {} {\  (\bibinfo {year} {2021})},\ \Eprint {http://arxiv.org/abs/2109.00193}
  {arXiv:2109.00193 [nucl-th]} \BibitemShut {NoStop}%
\bibitem [{\citenamefont {Klos}\ \emph {et~al.}(2013)\citenamefont {Klos},
  \citenamefont {Men\'endez}, \citenamefont {Gazit},\ and\ \citenamefont
  {Schwenk}}]{Klos:2013rwa}%
  \BibitemOpen
  \bibfield  {author} {\bibinfo {author} {\bibfnamefont {P.}~\bibnamefont
  {Klos}}, \bibinfo {author} {\bibfnamefont {J.}~\bibnamefont {Men\'endez}},
  \bibinfo {author} {\bibfnamefont {D.}~\bibnamefont {Gazit}}, \ and\ \bibinfo
  {author} {\bibfnamefont {A.}~\bibnamefont {Schwenk}},\ }\href {\doibase
  10.1103/PhysRevD.88.083516} {\bibfield  {journal} {\bibinfo  {journal} {Phys.
  Rev. D}\ }\textbf {\bibinfo {volume} {88}},\ \bibinfo {pages} {083516}
  (\bibinfo {year} {2013})},\ \bibinfo {note} {[Erratum: Phys.Rev.D 89, 029901
  (2014)]},\ \Eprint {http://arxiv.org/abs/1304.7684} {arXiv:1304.7684
  [nucl-th]} \BibitemShut {NoStop}%
\bibitem [{\citenamefont {Wang}\ \emph {et~al.}(2021)\citenamefont {Wang},
  \citenamefont {Yang},\ and\ \citenamefont {Zhu}}]{Wang:2021nbf}%
  \BibitemOpen
  \bibfield  {author} {\bibinfo {author} {\bibfnamefont {W.}~\bibnamefont
  {Wang}}, \bibinfo {author} {\bibfnamefont {W.-N.}\ \bibnamefont {Yang}}, \
  and\ \bibinfo {author} {\bibfnamefont {B.}~\bibnamefont {Zhu}},\ }\href@noop
  {} {\  (\bibinfo {year} {2021})},\ \Eprint {http://arxiv.org/abs/2111.04000}
  {arXiv:2111.04000 [hep-ph]} \BibitemShut {NoStop}%
\bibitem [{\citenamefont {Bednyakov}\ and\ \citenamefont
  {Simkovic}(2005)}]{Bednyakov:2004xq}%
  \BibitemOpen
  \bibfield  {author} {\bibinfo {author} {\bibfnamefont {V.~A.}\ \bibnamefont
  {Bednyakov}}\ and\ \bibinfo {author} {\bibfnamefont {F.}~\bibnamefont
  {Simkovic}},\ }\href@noop {} {\bibfield  {journal} {\bibinfo  {journal}
  {Phys. Part. Nucl.}\ }\textbf {\bibinfo {volume} {36}},\ \bibinfo {pages}
  {131} (\bibinfo {year} {2005})},\ \Eprint
  {http://arxiv.org/abs/hep-ph/0406218} {arXiv:hep-ph/0406218} \BibitemShut
  {NoStop}%
\bibitem [{\citenamefont {Bednyakov}\ and\ \citenamefont
  {Simkovic}(2006)}]{Bednyakov:2006ux}%
  \BibitemOpen
  \bibfield  {author} {\bibinfo {author} {\bibfnamefont {V.~A.}\ \bibnamefont
  {Bednyakov}}\ and\ \bibinfo {author} {\bibfnamefont {F.}~\bibnamefont
  {Simkovic}},\ }\href {\doibase 10.1134/S1063779606070057} {\bibfield
  {journal} {\bibinfo  {journal} {Phys. Part. Nucl.}\ }\textbf {\bibinfo
  {volume} {37}},\ \bibinfo {pages} {S106} (\bibinfo {year} {2006})},\ \Eprint
  {http://arxiv.org/abs/hep-ph/0608097} {arXiv:hep-ph/0608097} \BibitemShut
  {NoStop}%
\bibitem [{\citenamefont {Catena}\ \emph {et~al.}(2020)\citenamefont {Catena},
  \citenamefont {Emken}, \citenamefont {Spaldin},\ and\ \citenamefont
  {Tarantino}}]{Catena:2019gfa}%
  \BibitemOpen
  \bibfield  {author} {\bibinfo {author} {\bibfnamefont {R.}~\bibnamefont
  {Catena}}, \bibinfo {author} {\bibfnamefont {T.}~\bibnamefont {Emken}},
  \bibinfo {author} {\bibfnamefont {N.~A.}\ \bibnamefont {Spaldin}}, \ and\
  \bibinfo {author} {\bibfnamefont {W.}~\bibnamefont {Tarantino}},\ }\href
  {\doibase 10.1103/PhysRevResearch.2.033195} {\bibfield  {journal} {\bibinfo
  {journal} {Phys. Rev. Res.}\ }\textbf {\bibinfo {volume} {2}},\ \bibinfo
  {pages} {033195} (\bibinfo {year} {2020})},\ \Eprint
  {http://arxiv.org/abs/1912.08204} {arXiv:1912.08204 [hep-ph]} \BibitemShut
  {NoStop}%
\bibitem [{\citenamefont {Roberts}\ \emph {et~al.}(2016)\citenamefont
  {Roberts}, \citenamefont {Dzuba}, \citenamefont {Flambaum}, \citenamefont
  {Pospelov},\ and\ \citenamefont {Stadnik}}]{Roberts:2016xfw}%
  \BibitemOpen
  \bibfield  {author} {\bibinfo {author} {\bibfnamefont {B.~M.}\ \bibnamefont
  {Roberts}}, \bibinfo {author} {\bibfnamefont {V.~A.}\ \bibnamefont {Dzuba}},
  \bibinfo {author} {\bibfnamefont {V.~V.}\ \bibnamefont {Flambaum}}, \bibinfo
  {author} {\bibfnamefont {M.}~\bibnamefont {Pospelov}}, \ and\ \bibinfo
  {author} {\bibfnamefont {Y.~V.}\ \bibnamefont {Stadnik}},\ }\href {\doibase
  10.1103/PhysRevD.93.115037} {\bibfield  {journal} {\bibinfo  {journal} {Phys.
  Rev. D}\ }\textbf {\bibinfo {volume} {93}},\ \bibinfo {pages} {115037}
  (\bibinfo {year} {2016})},\ \Eprint {http://arxiv.org/abs/1604.04559}
  {arXiv:1604.04559 [hep-ph]} \BibitemShut {NoStop}%
\bibitem [{\citenamefont {Tan}\ \emph {et~al.}(2021)\citenamefont {Tan},
  \citenamefont {Derevianko}, \citenamefont {Dzuba},\ and\ \citenamefont
  {Flambaum}}]{Tan:2021nif}%
  \BibitemOpen
  \bibfield  {author} {\bibinfo {author} {\bibfnamefont {H.~B.~T.}\
  \bibnamefont {Tan}}, \bibinfo {author} {\bibfnamefont {A.}~\bibnamefont
  {Derevianko}}, \bibinfo {author} {\bibfnamefont {V.~A.}\ \bibnamefont
  {Dzuba}}, \ and\ \bibinfo {author} {\bibfnamefont {V.~V.}\ \bibnamefont
  {Flambaum}},\ }\href {\doibase 10.1103/PhysRevLett.127.081301} {\bibfield
  {journal} {\bibinfo  {journal} {Phys. Rev. Lett.}\ }\textbf {\bibinfo
  {volume} {127}},\ \bibinfo {pages} {081301} (\bibinfo {year} {2021})},\
  \Eprint {http://arxiv.org/abs/2105.08296} {arXiv:2105.08296 [hep-ph]}
  \BibitemShut {NoStop}%
\bibitem [{\citenamefont {Smith}\ \emph {et~al.}(2007)\citenamefont {Smith}
  \emph {et~al.}}]{Smith:2006ym}%
  \BibitemOpen
  \bibfield  {author} {\bibinfo {author} {\bibfnamefont {M.~C.}\ \bibnamefont
  {Smith}} \emph {et~al.},\ }\href {\doibase 10.1111/j.1365-2966.2007.11964.x}
  {\bibfield  {journal} {\bibinfo  {journal} {Mon. Not. Roy. Astron. Soc.}\
  }\textbf {\bibinfo {volume} {379}},\ \bibinfo {pages} {755} (\bibinfo {year}
  {2007})},\ \Eprint {http://arxiv.org/abs/astro-ph/0611671}
  {arXiv:astro-ph/0611671} \BibitemShut {NoStop}%
\bibitem [{\citenamefont {Dehnen}\ and\ \citenamefont
  {Binney}(1998)}]{Dehnen:1997cq}%
  \BibitemOpen
  \bibfield  {author} {\bibinfo {author} {\bibfnamefont {W.}~\bibnamefont
  {Dehnen}}\ and\ \bibinfo {author} {\bibfnamefont {J.}~\bibnamefont
  {Binney}},\ }\href {\doibase 10.1046/j.1365-8711.1998.01600.x} {\bibfield
  {journal} {\bibinfo  {journal} {Mon. Not. Roy. Astron. Soc.}\ }\textbf
  {\bibinfo {volume} {298}},\ \bibinfo {pages} {387} (\bibinfo {year}
  {1998})},\ \Eprint {http://arxiv.org/abs/astro-ph/9710077}
  {arXiv:astro-ph/9710077} \BibitemShut {NoStop}%
\bibitem [{\citenamefont {Aprile}\ \emph {et~al.}(2012)\citenamefont {Aprile}
  \emph {et~al.}}]{XENON100:2011cza}%
  \BibitemOpen
  \bibfield  {author} {\bibinfo {author} {\bibfnamefont {E.}~\bibnamefont
  {Aprile}} \emph {et~al.} (\bibinfo {collaboration} {XENON100}),\ }\href
  {\doibase 10.1016/j.astropartphys.2012.01.003} {\bibfield  {journal}
  {\bibinfo  {journal} {Astropart. Phys.}\ }\textbf {\bibinfo {volume} {35}},\
  \bibinfo {pages} {573} (\bibinfo {year} {2012})},\ \Eprint
  {http://arxiv.org/abs/1107.2155} {arXiv:1107.2155 [astro-ph.IM]} \BibitemShut
  {NoStop}%
\bibitem [{\citenamefont {Aprile}\ \emph
  {et~al.}(2019{\natexlab{b}})\citenamefont {Aprile} \emph
  {et~al.}}]{XENON:2019gfn}%
  \BibitemOpen
  \bibfield  {author} {\bibinfo {author} {\bibfnamefont {E.}~\bibnamefont
  {Aprile}} \emph {et~al.} (\bibinfo {collaboration} {XENON}),\ }\href
  {\doibase 10.1103/PhysRevLett.123.251801} {\bibfield  {journal} {\bibinfo
  {journal} {Phys. Rev. Lett.}\ }\textbf {\bibinfo {volume} {123}},\ \bibinfo
  {pages} {251801} (\bibinfo {year} {2019}{\natexlab{b}})},\ \Eprint
  {http://arxiv.org/abs/1907.11485} {arXiv:1907.11485 [hep-ex]} \BibitemShut
  {NoStop}%
\bibitem [{\citenamefont {Aprile}\ \emph {et~al.}(2016)\citenamefont {Aprile}
  \emph {et~al.}}]{XENON:2016jmt}%
  \BibitemOpen
  \bibfield  {author} {\bibinfo {author} {\bibfnamefont {E.}~\bibnamefont
  {Aprile}} \emph {et~al.} (\bibinfo {collaboration} {XENON}),\ }\href
  {\doibase 10.1103/PhysRevD.94.092001} {\bibfield  {journal} {\bibinfo
  {journal} {Phys. Rev. D}\ }\textbf {\bibinfo {volume} {94}},\ \bibinfo
  {pages} {092001} (\bibinfo {year} {2016})},\ \bibinfo {note} {[Erratum:
  Phys.Rev.D 95, 059901 (2017)]},\ \Eprint {http://arxiv.org/abs/1605.06262}
  {arXiv:1605.06262 [astro-ph.CO]} \BibitemShut {NoStop}%
\bibitem [{\citenamefont {Xia}\ \emph {et~al.}(2021)\citenamefont {Xia},
  \citenamefont {Xu},\ and\ \citenamefont {Zhou}}]{Xia:2021vbz}%
  \BibitemOpen
  \bibfield  {author} {\bibinfo {author} {\bibfnamefont {C.}~\bibnamefont
  {Xia}}, \bibinfo {author} {\bibfnamefont {Y.-H.}\ \bibnamefont {Xu}}, \ and\
  \bibinfo {author} {\bibfnamefont {Y.-F.}\ \bibnamefont {Zhou}},\ }\href@noop
  {} {\  (\bibinfo {year} {2021})},\ \Eprint {http://arxiv.org/abs/2111.05559}
  {arXiv:2111.05559 [hep-ph]} \BibitemShut {NoStop}%
\end{thebibliography}%
 \end{document}